\documentclass[12pt, a4paper]{article}
 \usepackage[font=small,format=plain,labelfont=bf,up,textfont=normal,up,justification=justified,singlelinecheck=false]{caption}
\usepackage{subcaption}
\usepackage{mathrsfs}
\usepackage[a4paper, left=2cm,right=2cm]{geometry}
\usepackage[colorlinks=true,linkcolor=black,citecolor=teal,urlcolor=MidnightBlue,filecolor=black]{hyperref}
\usepackage{amsfonts}
\usepackage{amsmath,amssymb}
\usepackage{setspace}
\usepackage{slashed}
\usepackage{braket}

\usepackage[dvipsnames]{xcolor}
\definecolor{SchoolColor}{rgb}{0.6471, 0.1098, 0.1882} 
\usepackage{subfloat}

\usepackage[utf8,applemac]{inputenc}
\usepackage{tensor}
\usepackage{cite}
\usepackage{tikz}
\usetikzlibrary{calc}
\usetikzlibrary{patterns}
\usetikzlibrary{arrows.meta}

\usepackage{graphicx}
\usepackage{bm} 
\allowdisplaybreaks[4]
\usepackage[utf8,applemac]{inputenc}
\usepackage{tensor}
\usepackage{cite}
\usepackage{tikz}
\usepackage{graphicx}
\graphicspath{{figure/}}
\usepackage{array}
\usepackage{booktabs}

\usepackage{multirow}

\usepackage{dcolumn}
\usepackage{bm}

\usepackage{verbatim}

\usepackage{textcomp} 
\usepackage{graphicx} 

\numberwithin{equation}{section}
\newcommand{\bea}{\begin{eqnarray}}
\newcommand{\eea}{\end{eqnarray}}
\newcommand{\be}{\begin{equation}}
\newcommand{\ee}{\end{equation}}
\newcommand{\bs}{\begin{subequations}}
\newcommand{\es}{\end{subequations}}
\def\nn{\nonumber}

\newcommand{\beqs}{\begin{eqnarray}}
\newcommand{\eeqs}{\end{eqnarray}}
\numberwithin{equation}{section}

\newcommand{\Rmnum}[1]{\uppercase\expandafter{\romannumeral #1\relax}}

\def\c.c.{\mathrm{c.c.}}

\def\a{\alpha}

\newcommand{\xhy}{\color{green}}
\newcommand{\yang}{\color{orange}}

\begin{document}
\begin{titlepage}

\begin{flushright}\vspace{-3cm}
{\small
\today }\end{flushright}
\vspace{0.5cm}
\begin{center}
	{{ \LARGE{\bf{On the definition of Carrollian  amplitudes  \vspace{8pt}\\ in general dimensions }}}}\vspace{5mm}

	\centerline{\large{Wen-Bin Liu\footnote{liuwenbin0036@hust.edu.cn}, Jiang Long\footnote{longjiang@hust.edu.cn}, Hong-Yang Xiao\footnote{xiaohongyang@hust.edu.cn} \& Jing-Long Yang\footnote{yangjinglong@hust.edu.cn}}}
	\vspace{2mm}
	\normalsize
	\bigskip\medskip

	\textit{School of Physics, Huazhong University of Science and Technology, \\ Luoyu Road 1037, Wuhan, Hubei 430074, China
	}
	
	\vspace{25mm}
	
	\begin{abstract}
		\noindent
		{Carrollian amplitude is the natural object that defines the correlator of the boundary Carrollian field theory. In this work, we will elaborate on its proper definition in general dimensions. We use the vielbein field on the unit sphere to define the fundamental field with non-vanishing helicity in the local Cartesian frame which is the building block of the Carrollian amplitude. In general dimensions, the Carrollian amplitude is related to the momentum space scattering matrix by a modified Fourier transform. The Poincar\'e transformation law  of the Carrollian amplitude in this definition has been discussed. We also find an isomorphism between the local rotation of the vielbein field and the superduality transformation. }\end{abstract}
	

\end{center}

\end{titlepage}
\tableofcontents

\section{Introduction}
Flat holography is stimulated by the success of AdS/CFT and has attracted much attention due to its potential applications to realistic physical processes.  Early efforts on the holographic principle \cite{1993gr.qc....10026T,Susskind:1994vu} in asymptotically flat spacetime can be traced back to the work of \cite{Polchinski:1999ry,Susskind:1998vk,Giddings:1999jq,deBoer:2003vf,Arcioni:2003xx,Arcioni:2003td,Mann:2005yr}. Up to now, there are two main approaches to tackle this interesting problem.
The first method is called celestial holography \cite{Pasterski:2016qvg,Pasterski:2017kqt,Pasterski:2017ylz}.  In this approach, the dual field theory is assumed to live on the two-dimensional celestial sphere. In the second approach, the dual field theory is assumed to live on the three-dimensional null boundary of the asymptotically  flat spacetime \cite{Donnay:2022aba,Bagchi:2022emh}. This method is called Carrollian holography since the  null boundary is actually a Carrollian manifold \cite{Une, Gupta1966OnAA}. Despite their extraordinary unlikeness, the two approaches are actually equivalent to each other at the level of scattering amplitude. In celestial holography, the usual massless S matrix is mapped  to the so-called celestial amplitude by a Mellin transform. On the other hand, the massless S matrix is mapped to the Carrollian amplitude by a Fourier transform. The scattering amplitude, celestial amplitude and Carrollian amplitude form an equivalent triangle by these integral transforms \cite{Donnay:2022wvx}. Despite various developments on celestial amplitude, the work on Carrollian amplitude is relatively fewer. The two-point Carrollian amplitudes have been derived in \cite{Liu:2022mne, Donnay:2022wvx} and the three-point Carrollian amplitudes can also be found in \cite{Salzer:2023jqv,Nguyen:2023miw}. The tree-level Carrollian amplitudes for gluons and gravitons have been explored systematically \cite{Mason:2023mti}. The Feynman rules and loop structure of Carrollian amplitude have been clarified in \cite{Liu:2024nfc}. The Carrollian amplitude in the context of string theory can be found in \cite{Stieberger:2024shv}. Recently, the potential connection to eikonal amplitude has been mentioned in \cite{Adamo:2024mqn} and the relation between the AdS limit of Witten diagrams \cite{Witten:1998qj} and Feynman diagrams has been investigated \cite{Alday:2024yyj}.  The differential equation of Carrollian amplitudes has also been studied \cite{Ruzziconi:2024zkr} and connected to the BG equation in celestial MHV gluon amplitude \cite{Banerjee:2020vnt}.  

The Carrollian holography has many interesting features that deserve further study. At first, the dual theory is based on the Carrollian manifold which is a natural geometric object at the boundary of asymptotically flat spacetime. The BMS group \cite{Bondi:1962px,Sachs:1962wk} and its various extensions \cite{Barnich:2010eb,Campiglia:2014yka, Campiglia:2015yka}, the asymptotic symmetry group of asymptotically flat spacetime is a geometric symmetry of the boundary manifold \cite{Duval:2014uoa,Duval:2014uva,Duval:2014lpa}. Second, this approach matches perfectly with the asymptotic quantization  \cite{1978JMP....19.1542A, Ashtekar:1981bq, Ashtekar:1981sf, Ashtekar:1987tt}, which has been shown by a series of papers \cite{Liu:2022mne,Liu:2023qtr,Liu:2023gwa,Li:2023xrr,Liu:2023jnc,Liu:2024nkc}. One can find more comparisons between celestial holography and Carrollian holography in \cite{Bagchi:2023cen}.

However, there are still various open problems on the Carrollian amplitude. One of the crucial questions is how to define spinning Carrollian amplitude from bulk reduction. Previous work \cite{Donnay:2022wvx}  on Carrollian amplitudes chose to use stereographic coordinates of the celestial sphere. For gluons and gravitons, the Carrollian amplitude is computed with definite helicity \cite{Mason:2023mti}. While this may have the advantage of adopting the technology of 2d CFT and the spinor helicity formalism in four dimensions \cite{Xu:1986xb}, there are no similar stereographic coordinates in higher dimensions and the details of the spinor helicity formalism depend on the dimensions\cite{Cheung:2009dc,Caron-Huot:2010nes,Boels:2012ie}. 
Moreover, it is well known that there are tensions on the asymptotic symmetry group in higher dimensions \cite{Hollands:2003ie, Hollands:2004ac,Tanabe:2011es} and soft theorems \cite{Kapec:2015vwa}. It is interesting to explore the extension of Carrollian holography to higher dimensions.
In principle, Carrollian amplitude is a geometric quantity whose definition should be independent of the coordinate choice of the celestial sphere. Since one can define scattering amplitude in general dimensions, there should be a corresponding Carrollian amplitude once flat holography is valid. In this paper, we will try to solve these problems  in the framework of bulk reduction \cite{Liu:2022mne,Liu:2024nkc}. In this formalism, the boundary field is obtained by asymptotic expansion of the bulk field $\tt{f}$ and there is always a fundamental field $F$ 
that represents the radiative  degree of freedom at the leading order of the fluctuation.
The canonical quantization of the fundamental field inherits the canonical quantization of the bulk field and one can define the asymptotic state in Carrollian space straightforwardly. It turns out that the Carrollian space state is related to the momentum space states by an integral transform which can be used to define the Carrollian amplitude. At the boundary, the Carrollian amplitude is reinterpreted as the correlator of the fundamental fields. One may use the vielbein field on the unit sphere to switch the spinning fundamental field into the local Cartesian frame. The corresponding Carrollian amplitude has nice properties under Poincar\'e transformation. The result can be extended to general dimensions and the integral transform should be modified compared to four dimensions. Interestingly, the ambiguity of the vielbein field in the Cartesian frame matches exactly with the superduality transformation.

This paper is organized as follows. In section \ref{pre}, we review the boundary Poincar\'e transformation in the framework of bulk reduction. In section \ref{poin}, we will define the vector Carrollian amplitude in four dimensions. In the following section,  the results are extended in gravitons. In section \ref{hd}, we define Carrollian amplitude in higher dimensions. In the following section, we will discuss the isomorphism between the local rotation of the vielbein field and the superduality transformation.  We will conclude in section \ref{conc}. Technical details are  relegated to several appendices.

\section{Preliminaries}\label{pre}
In \cite{Liu:2024nfc}, the finite Poincar\'e transformation has been studied carefully by reducing the bulk transformation to the null boundary. We will use Greek alphabet  $\mu,\nu,\cdots$ to denote Cartesian coordinates and $\alpha,\beta,\cdots$ to denote retarded or advanced coordinates in Minkowski spacetime $\mathbb{R}^{1,3}$. The Latin alphabet $i,j,k,\cdots$ are used to label spatial directions in Cartesian coordinates. The capital Latin alphabet  $A,B,C,\cdots$ are the labels of spherical components.  Therefore, the metric of the four-dimensional Minkowski spacetime $\mathbb{R}^{1,3}$ can be written in Cartesian coordinates as 
\bea 
ds^2=g_{\mu\nu}dx^\mu dx^\nu=-dt^2+dx^i dx^i,\quad \mu,\nu=0,1,2,3
\eea 
where the Minkowski matrix is $g_{\mu\nu}=\text{diag}(-1,+1,+1,+1)$. Switching to the spherical coordinates $(t,r,\theta^A),\ A=1,2$, the metric becomes 
\bea 
ds^2=-dt^2+dr^2+r^2(d\theta^2+\sin^2\theta d\phi^2).
\eea To simplify notation, we will also use the label $\Omega$ to collect the spherical angles and the boldface Latin alphabet $\bm x$ to denote spatial vectors. The metric of the unit sphere $S^2$ is denoted as $\gamma_{AB}$. One can embed the sphere into {a} flat space and the unit normal vector is 
\be 
\bm n=\frac{\bm x}{r}=(\sin\theta\cos\phi,\sin\theta\sin\phi,\cos\theta).
\ee 
We may pullback the intrinsic metric $\gamma_{AB}$ to the three-dimensional space 
\bea 
\gamma_{ij}=\delta_{ij}-n_i n_j,
\eea where $\delta_{ij}$ is the Kronecker delta.
We can also define two null vectors $n^\mu$ and $\bar n^\mu$ as 
\bea 
n^\mu=(1,n^i),\quad \bar n^\mu=(-1,n^i),
\eea and the four-dimensional version of the metric $\gamma_{AB}$ is 
\bea 
\gamma_{\mu\nu}=g_{\mu\nu}-\frac{1}{2}(n_\mu \bar{n}_\nu+n_\nu\bar n_\mu).
\eea 

The retarded time is defined as $u=t-r$ and the metric of the Minkowski spacetime can be written in retarded coordinates $(u,r,\Omega)$
\be 
ds^2=-du^2-2du dr+r^2(d\theta^2+\sin^2\theta d\phi^2).
\ee 
The future null infinity $\mathcal{I}^+$ is parameterized by three coordinates $(u,\Omega)$. The Cartesian coordinates may be written in terms of retarded coordinates as 
\be 
x^\mu=u \bar{m}^\mu+r n^\mu,\quad \bar{m}^\mu=\frac{1}{2}(n^\mu-\bar n^\mu)=(1,0,0,0).
\ee Therefore, the Jacobian from Cartesian coordinates to retarded coordinates is 
\bea 
\frac{\partial x^\mu}{\partial x'^{\alpha}}=\bar m^\mu\delta_\a^u+n^\mu\delta_\a^r-rY^\mu_A\delta_\a^A,\label{Jacobian}
\eea where we have defined 
\be 
Y^\mu_A=-\nabla_A n^\mu.
\ee 
Similarly, one may define an advanced time $v=t+r$ and the metric can be written in advanced coordinates $(v,r,\Omega)$
\be ds^2=-dv^2+2dv dr+r^2(d\theta^2+\sin^2\theta d\phi^2).
\ee The past null infinity $\mathcal{I}^-$ is parameterized by the coordinates $(v,\Omega)$. The Cartesian coordinates may be written in terms of advanced coordinates as 
\be 
x^\mu=v \bar{m}^\mu+r \bar n^\mu.
\ee 
 
As has been shown in \cite{Liu:2024nfc}, the spacetime translation 
\be 
x'^\mu=x^\mu+c^\mu
\ee parameterized by a constant vector $c^\mu$ can be written in terms of boundary coordinates
\bea 
u'=u-c\cdot n,\quad \Omega'=\Omega.\label{stOmega}
\eea 
For the Lorentz transformation 
\be
x'^\mu=\Lambda^\mu_{\ \nu}x^\nu,\quad \Lambda^\mu_{\ \nu}\Lambda^\rho_{\ \sigma}g_{\mu\rho}=g_{\nu\sigma},
\ee  the corresponding boundary coordinate transformation is 
\bea 
u'=\Gamma^{-1}u,\quad n'^i=\Gamma^{-1}\Gamma^i,\label{Lorentztansf}
\eea where 
\bea 
\Gamma=\Lambda^0_{\ \nu}n^\nu,\quad \Gamma^i=\Lambda^i_{\ \nu}n^\nu.
\eea 

\paragraph{Boundary metric.}  It is understood that Minkowski matrix is invariant under general Poincar\'e transformations. Since the spatial radius is invariant under spacetime translation near $\mathcal{I}^+$
\be 
r'=r+\mathcal{O}(1),
\ee the boundary metric $\gamma_{AB}$ is invariant under spacetime translation 
\be 
\gamma_{AB}(\Omega)=\gamma'_{AB}(\Omega')=\gamma'_{AB}(\Omega).
\ee At the last step, we have used the transformation law \eqref{stOmega}. However, since the spatial radius $r$ is rescaled under Lorentz transformation near $\mathcal{I}^+$
\be r'=\Gamma\ r+\mathcal{O}(1),\label{rscale}
\ee we may read out the transformation law for the boundary metric 
\bea 
\gamma'_{AB}(\Omega')=\lim_{r'\to\infty}r'^{-2}g'_{AB}(u',r',\Omega')=\lim_{r\to\infty}r^{-2}\Gamma^{-2}\frac{\partial x^\mu}{\partial \theta'^A}\frac{\partial x^\nu}{\partial\theta'^B}g_{\mu\nu}(x).
\eea Note the Jacobian \eqref{Jacobian}, we find the following Lorentz transformation law
\bs\begin{align} 
\gamma'_{AB}(\Omega')&=\Gamma^{-2}\frac{\partial\theta^C}{\partial\theta'^A}\frac{\partial\theta^D}{\partial\theta'^B}\gamma_{CD}(\Omega),\label{gamma}\\
\gamma'^{AB}(\Omega')&=\Gamma^2\frac{\partial\theta^A}{\partial\theta'^C}\frac{\partial\theta^B}{\partial\theta'^D}\gamma^{CD}(\Omega).\label{gammainverse}
\end{align}\es This transformation law is distinguished from the intrinsic coordinate transformation on $S^2$ for the metric. This fact has been emphasized in \cite{Liu:2024nfc}, and we will elaborate it later.

\paragraph{Scalar field.} Under a general bulk diffeomorphism, the bulk scalar field $\Phi(x)$ obeys the following transformation law 
\be 
\Phi'(x')=\Phi(x).
\ee 
Combining with the fall-off condition 
\bea 
\Phi(x)=\frac{\Sigma(u,\Omega)}{r}+\cdots,
\eea the radiative field $\Sigma(u,\Omega)$ would be regarded as a boundary field whose transformation law is 
\be 
\Sigma'(u',\Omega')=\Sigma(u,\Omega)
\ee for spacetime translation and 
\bea 
\Sigma'(u',\Omega')=\Gamma\ \Sigma(u,\Omega)\label{transformationSigma}
\eea for Lorentz transformation.

\section{Vector field}\label{poin}
In this section, we will work on the Carrollian amplitude for the vector field theory in four dimensions. At first, we will discuss the Poincar\'e transformation law of the boundary field. We conclude that it is necessary to introduce a vielbein field to define a boundary field $A_a$ in the local Cartesian frame in order to obtain a concise transformation law. It turns out that it is convenient to define Carrollian amplitude using this boundary field in the local frame. The corresponding Carrollian amplitude is mapped to the standard scattering amplitude in spinor helicity formalism by a Fourier transform. The method will be generalized to the gravitational theory and higher dimensions in the following sections.
\subsection{Transformation law}
For a vector field $a_\mu$, under a general Poincar\'e transformation 
\be 
x'^\mu=\Lambda^\mu_{\ \nu}x^\nu+c^\mu
\ee it transforms as
\be 
a_\mu'(x')=\frac{\partial x^\nu}{\partial x'^\mu}a_\nu(x)=\left(\Lambda^{-1}\right)^\nu_{\ \mu} a_\nu(x)=\Lambda_\mu^{\ \nu}a_\nu(x).
\ee 
For a massless vector field, to extract the radiative degree of freedom, we may impose the fall-off condition \cite{jackson2021classical} 
\be 
a_\mu(x)=\frac{A_{\mu}(u,\Omega)}{r}+\cdots.\label{falloffa}
\ee 


For spacetime translation, the spatial radius $r'$ is invariant in the sense 
\be 
r'=r+\mathcal{O}(1),
\ee therefore, the field $A_\mu(u,\Omega)$ is invariant under spacetime translation 
\be 
A'_\mu(u',\Omega')=A_\mu(u,\Omega)\quad \text{for spacetime translation}.
\ee On the other hand, since the spatial radius $r'$ is rescaled by the factor $\Gamma$ under Lorentz transformation, 
 the field $A_\mu(u,\Omega)$ transforms as 
\be 
A'_\mu(u',\Omega')=\Gamma\Lambda_\mu^{\ \nu} A_\nu(u,\Omega)\quad\text{for Lorentz transformation}.
\ee 
The radiative field $A_A(u,\Omega)$ is defined as  \cite{Liu:2023qtr}
\be 
A_A(u,\Omega)=-Y^\mu_A(\Omega)A_\mu(u,\Omega).
\ee Note that the transformation of the null vector 
\bea 
n'^\mu=(1,n'^i)=(1,\Gamma^{-1}\Gamma^i)=\Gamma^{-1}\Lambda^\mu_{\ \nu}n^\nu,
\eea we find 
\bea 
Y'^{\mu}_{A}(\Omega')&=&-\nabla'_An'^\mu=-\frac{\partial\theta^B}{\partial\theta'^A}\nabla_B(\Gamma^{-1}\Lambda^\mu_{\ \nu}n^\nu)\nn\\&=&-\frac{\partial\theta^B}{\partial\theta'^A}\Gamma^{-1}\Lambda^\mu_{\ \nu}\nabla_B n^\nu+\Gamma^{-1}\frac{\partial\theta^B}{\partial\theta'^A}\left(\nabla_B\Gamma\right) \Gamma^{-1}\Lambda^\mu_{\ \nu}n^\nu\nn\\&=&\Gamma^{-1}\frac{\partial\theta^B}{\partial\theta'^A}\Lambda^{\mu}_{\ \nu}\left(Y^\nu_B(\Omega)+ n^\nu\nabla_B\log\Gamma\right).\label{Y}
\eea Therefore, 
\bea 
A'_A(u',\Omega')&=&-Y'^{\mu}_{A}(\Omega')A'_\mu(u',\Omega')=-\Gamma^{-1}\frac{\partial\theta^B}{\partial\theta'^A}\Lambda^{\mu}_{\ \nu}\left(Y^\nu_B(\Omega)+ n^\nu\nabla_B\log\Gamma\right)\Gamma \Lambda_\mu^{\ \rho}A_\rho(u,\Omega).\nn\\
\eea The $\Gamma$ factors are canceled except for the $\log\Gamma$. At the same time, the Lorentz transformation matrices $\Lambda^\mu_{\ \nu}$ are canceled due to the property $\Lambda^\mu_{\ \nu}\Lambda_\mu^{\ \rho}=\delta^\rho_\nu$. 
\bea 
A'_A(u',\Omega')=\frac{\partial\theta^B}{\partial\theta'^A}A_B(u,\Omega)-n^\nu A_\nu(u,\Omega)\frac{\partial\theta^B}{\partial\theta'^A}\nabla_B\log\Gamma.\label{AAtrans}
\eea 
To eliminate the second term, one may impose a further fall-off condition 
\be 
a_r=\mathcal{O}(r^{-2})\quad\Rightarrow\quad n^\nu A_\nu(u,\Omega)=0
\ee besides \eqref{falloffa}.
This is satisfied in the radial gauge $a_r=0$ which is chosen in \cite{Liu:2023qtr}.
Note that $A_r=0$ can also be checked in Lorenz gauge since $n^\mu$ is proportional to the four-momentum near $\mathcal{I}^+$. Therefore, we obtain the following transformation law for the radiative field under Lorentz transformations
\bea 
A'_A(u',\Omega')=\frac{\partial\theta^B}{\partial\theta'^A}A_B(u,\Omega).\label{AA}
\eea Combining with the \eqref{gammainverse}, we find 
\bea 
A'^A(u',\Omega')=\Gamma^2 \frac{\partial\theta'^A}{\partial\theta^B}A^B(u,\Omega).\label{AuA}
\eea Comparing \eqref{AA} with \eqref{AuA}, the radiative field $A_A(u,\Omega)$ is not a standard vector field on $S^2$. To fix this point, we introduce a vielbein field $e_A^a$ on $S^2$ which satisfies  the conditions 
\bea 
e_A^a e_B^b \gamma^{AB}=\gamma^{ab},\quad e_A^a e_B^b \gamma_{ab}=\gamma_{AB},\quad a=1,2,\label{conditions}
\eea where $\gamma_{ab}$ is the flat metric in the local Cartesian frame. The inverse of the vielbein field is $e^B_a$ whose indices are raised or lowered by $\gamma^{AB}$ or $\gamma_{ab}$, respectively. The flat metric $\gamma_{ab}$ is invariant under Lorentz transformation, while the metric $\gamma_{AB}$ is transformed as \eqref{gamma}. Therefore, the vielbein field may  transform as follows 
\bs\label{LorentzR}\label{v}\begin{align}
e_A'^{a}(\Omega')&=\Gamma^{-1}\frac{\partial\theta^B}{\partial\theta'^A}R^a_{\ b}(\Omega)e_B^b(\Omega),\\
e'^{A}_a(\Omega')&=\Gamma \frac{\partial\theta'^A}{\partial\theta^B}R_a^{\ b}(\Omega)e^B_{\ b}(\Omega).
\end{align}\es We have inserted a $2\times 2$ matrix $R^a_{\ b}(\Omega)$ into the transformation law since the constraints \eqref{conditions} only fix the vielbein field up to a local rotation.
The rotation matrix $R^a_{\ b}$ is a local orthogonal matrix 
\be 
R^a_{\ b}R^c_{\ d}\gamma_{ac}=\gamma_{bd}.
\ee 
To fix the matrix $R^a_{\ b}$, we should impose more constraints. A natural choice is that the form of the vielbein field is unchanged  under the local Lorentz transformation 
\be 
e'^a_A(\Omega')=e_A^a(\Omega').\label{inve}
\ee We will leave the rotation matrix $R^a_{\ b}$ free at this moment and will discuss the above condition later. 
With the vielbein field, we may define the following vector field  in the local Cartesian frame 
\bea 
A_a(u,\Omega)=e^A_a(\Omega)A_A(u,\Omega),\quad A^a(u,\Omega)=e^a_A(\Omega)A^A(\Omega)
\eea which is invariant under spacetime translation 
\be 
A_a'(u',\Omega')=A_a(u,\Omega),\quad u'=u-c\cdot n,\quad \Omega'=\Omega.
\ee For the Lorentz transformation \eqref{Lorentztansf}, we find 
\bea 
A_a'(u',\Omega')=\Gamma R_a^{\ b}A_b(u,\Omega),\quad A'^{a}(u',\Omega')=\Gamma R^a_{\ b}A^b(u,\Omega).\label{LorentztransfAa}
\eea
The field $A_a$ in the Cartesian frame has much better properties than the original field $A_A$. Firstly, the fields with upper and lower index in the Cartesian frame transform in the same way under Lorentz transformations. Secondly, the transformation law \eqref{LorentztransfAa} shares  the same structure as the one for the scalar field \eqref{transformationSigma} up to a local rotation. 
The rotation matrix reflects the spin of the field $A_a$.
\subsection{Canonical quantization}
The canonical quantization for the radiative field $A_A(u,\Omega)$ at the boundary $\mathcal{I}^{\pm}$ has been worked out in \cite{Liu:2023qtr}. In this subsection, we will translate it to the form that is suitable to define Carrollian amplitude. 
By imposing the Lorenz gauge 
\be 
\partial_\mu a^\mu=0,\label{lorgauge}
\ee the electromagnetic field $a_\mu$ may be quantized   using annihilation and creation operators $b_{a,\bm p}, c^\dagger_{a,\bm p}$ 
\bea 
a_\mu(x)&=&\int \frac{d^3\bm p}{\sqrt{(2\pi)^3}}\frac{1}{\sqrt{2\omega_{\bm p}}}[\epsilon^{{a}}_\mu(\bm p)b_{{a},\bm p}e^{-i\omega t+i\bm p\cdot\bm x}+\epsilon^{{a}}_\mu(\bm p)c^\dagger_{{a},\bm p}e^{i\omega t-i\bm p\cdot\bm x}],\label{modeexpansion}
\eea  where the vector $\bm p$ is the momentum and $\omega$ is the energy of the corresponding mode. The operators $b_{a,\bm p},\ c_{a,\bm p}^\dagger$ satisfy the commutation relations 
\be 
[b_{a,\bm p},b_{b,\bm p'}]=[c^\dagger_{a,\bm p},c^\dagger_{b,\bm p'}]=0,\quad [b_{a,\bm p},c_{b,\bm p'}^\dagger]=\gamma_{ab}\delta(\bm p-\bm p').
\ee For a massless particle, we have
\be 
\omega=|\bm p|.
\ee The index ${a}$ denotes the polarization of the corresponding mode.  The polarization vector  $\epsilon_\mu^{{a}}(\bm p)$ has two physical degrees of freedom and is orthogonal to the four-momentum
\be 
\epsilon_\mu^a(\bm p)\cdot p^\mu=0.
\ee 
They also satisfy the orthogonality and completeness relations
\be 
\epsilon_\mu^a(\bm p) \gamma^{\mu\nu}\epsilon_\nu^b(\bm p)=\gamma^{ab},\quad \epsilon_{\mu}^{a}(\bm p)\gamma_{ab}\epsilon_\nu^{b}(\bm p)=\gamma_{\mu\nu}.\label{orthocomp}
\ee 
The Hermitian conjugate of the vector field is 
\bea 
\left(a_\mu(x)\right)^\dagger=\int \frac{d^3\bm p}{\sqrt{(2\pi)^3}}\frac{1}{\sqrt{2\omega_{\bm p}}}[\left(\epsilon^{{a}}_\mu(\bm p)\right)^*b_{{a},\bm p}^\dagger e^{i\omega t-i\bm p\cdot\bm x}+\left(\epsilon^{{a}}_\mu(\bm p)\right)^*c_{{a},\bm p}e^{-i\omega t+i\bm p\cdot\bm x}].
\eea Note that the complex conjugate of the polarization vector $\epsilon_\mu^a$ should be a linear superposition of the original ones
\be 
\left(\epsilon^{{a}}_\mu\right)^*=C^a_{\ b}\epsilon_\mu^b.\label{complexC}
\ee For the vector field to be real, we should impose the condition 
\bea 
C^a_{\ b}c_{a,\bm p}=b_{b,\bm p},\quad C^a_{\ b}b^\dagger_{a,\bm p}=c_{b,\bm p}^\dagger.
\eea 
It follows that 
\bea 
[b_{a,\bm p},b_{b,\bm p'}^\dagger]=C_{ba}\delta(\bm p-\bm p'),\quad [c_{a,\bm p},c_{b,\bm p'}^\dagger]=C_{ab}\delta(\bm p-\bm p')
\eea where 
\be 
C_{ab}=\gamma_{ac}C^c_{\ b}.
\ee Another consistency condition is 
\bea 
b_{a,\bm p}=C^b_{\ a}c_{b,\bm p}=C^b_{\ a}\left(C^{\dagger}\right)^{\ c}_bb_{c,\bm p}\quad\Rightarrow\quad C^* C=1.
\eea 
A convenient choice of $C_{ab}$ is 
\be 
C_{ab}=\delta_{ab}.\label{Cab}
\ee 

With the fall-off condition \eqref{falloffa}, we can read out the leading coefficients
\bs\begin{align}
A_\mu(u,\Omega)&=-\frac{i}{\sqrt{4\pi}}\int_0^\infty d\omega \sqrt{\omega} \epsilon_\mu^a(\bm p)(b_{a,\bm p}e^{-i\omega u}-c_{a,\bm p}^\dagger e^{i\omega u}),\\
A_\mu(v,\Omega)&=\frac{i}{\sqrt{4\pi}}\int_0^\infty d\omega \sqrt{\omega}\epsilon_\mu^a(\bm p^{\text{P}})(b_{a,\bm p^{\text{P}}}e^{-i\omega v}-c_{a,\bm p^{\text{P}}}^\dagger e^{i\omega v}).
\end{align}\es The notation $\bm p^{\text{P}}$ represents the image of the momentum under antipodal map,
\be 
v\to u,\quad \Omega\to \Omega^{\text{P}}.
\ee 

In spherical coordinates, $\bm p^{\text{P}}$ is
\be 
\bm p^{\text{P}}=(\omega,\pi-\theta,\pi+\phi).
\ee 
In general, we will use the notation ${q}^{\text{P}}$ to denote the image of a quantity $q$ under antipodal map. 
The field $A_\mu(v,\Omega)$ is mapped to\footnote{Similar to the scalar field, the antipodal map of the field is only defined up to a constant phase factor.}
\bea 
A^{\text{P}}_\mu(u,\Omega)=A_\mu(v\to u,\Omega\to \Omega^{\text{P}})=\frac{i}{\sqrt{4\pi}}\int_0^\infty d\omega \sqrt{\omega}\epsilon_\mu^a(\bm p)(b_{a,\bm p}e^{-i\omega u}-c^\dagger_{a,\bm p}e^{i\omega u})
\eea
As a vector field on $S^2$, the antipodal map will send $Y^\mu_A(\Omega)$ to 
\be 
\left(Y^{\text{P}}\right)^\mu_A(\Omega^{\text{P}})=\frac{\partial\theta^B}{\partial\theta'^A}Y^\mu_{B}(\Omega)=(-Y_\theta^\mu(\Omega),Y_\phi^\mu(\Omega)).
\ee Then the antipodal field $\left(Y^{\text{P}}\right)^\mu_A(\Omega)$ is 
\be
\left(Y^{\text{P}}\right)^\mu_A(\Omega)=(-Y^\mu_\theta(\Omega^{\text{P}}),Y^\mu_\phi(\Omega^{\text{P}}))=-Y^\mu_A(\Omega).
\ee 
As a consequence, we can find the fundamental field as 
\bs\begin{align}
    A_A(u,\Omega)&=-Y^\mu_{{A}}(\Omega)A_\mu(u,\Omega)=\frac{i}{\sqrt{4\pi}}\int_0^\infty d\omega \sqrt{\omega} Y^\mu_A(\Omega)\epsilon_\mu^a(\bm p)(b_{a,\bm p}e^{-i\omega u}-c_{a,\bm p}^\dagger e^{i\omega u}),\\
    A_A^{\text{P}}(u,\Omega)&=-\left(Y^{\text{P}}\right)^\mu_A(\Omega)
    A_\mu^{\text{P}}(u,\Omega)=\frac{i}{\sqrt{4\pi}}\int_0^\infty d\omega \sqrt{\omega} Y^\mu_A(\Omega)\epsilon_\mu^a(\bm p)(b_{a,\bm p}e^{-i\omega u}-c_{a,\bm p}^\dagger e^{i\omega u})
\end{align}\es 
Therefore, under antipodal map, the field $A^{\text{P}}_A(u,\Omega)$ is exactly the same as the form of $A_A(u,\Omega)$. Interestingly, there is a natural veilbein field on $S^2$
\be 
e^a_A(\Omega)=-Y^\mu_A(\Omega)\epsilon_\mu^a(\bm p).\label{veil}
\ee We have proved this point using orthogonality and completeness relations of the polarization vector in Appendix \ref{vielbeinapp}. We may define a boundary state located at $(u,\Omega)$ as follows 
\bea 
|A_A(u,\Omega)\rangle=A_A(u,\Omega)|0\rangle=\frac{i}{8\pi^2}\int_0^\infty d\omega e^{i\omega u}e_A^a(\Omega)|\bm p,a\rangle
\eea 
where the state $|\bm p,a\rangle$ with definite momentum $\bm p$ and polarization $a$ is 
\be 
|\bm p,a\rangle=\sqrt{(2\pi)^3 2\omega_{\bm p}}c_{a,\bm p}^\dagger|0\rangle.
\ee In terms of the field of the Cartesian frame, we find 
\bea 
|A_a(u,\Omega)\rangle=e^A_a(\Omega)A_A(u,\Omega)|0\rangle=\frac{i}{8\pi^2}\int_0^\infty d\omega e^{i\omega u}|\bm p,a\rangle.\label{Aapa}
\eea Inversely, 
\bea 
|\bm p,a\rangle=-4\pi i \int_{-\infty}^\infty du e^{-i\omega u}|A_a(u,\Omega)\rangle.\label{paAa}
\eea The relations \eqref{Aapa} and \eqref{paAa} are exactly the same as the ones for the scalar field, except  for the extra polarizations.  

Similarly, we find
\begin{align}
  |A^{\rm P}_A(u,\Omega)\rangle=A^{\rm P}_A(u,\Omega)|0\rangle&=\frac{i}{8\pi^2}\int_0^\infty d\omega e^{i\omega u}e_A^a(\Omega)|\bm p,a\rangle,
\end{align}
and thus
\begin{align}
  |A^{\rm P}_a(u,\Omega)\rangle=e^A_a(\Omega)A^{\rm P}_A(u,\Omega)|0\rangle=\frac{i}{8\pi^2}\int_0^\infty d\omega e^{i\omega u}|\bm p,a\rangle.\label{AaP}
\end{align}

We may also define the state 
\bea 
\langle A_A(u,\Omega)|=\langle 0|A_A(u,\Omega)=-\frac{i}{8\pi^2}\int_0^\infty d\omega e^{-i\omega u}e_A^a(\Omega)\langle a,\bm p|
\eea with the convention 
\bea 
\langle a,\bm p|=\langle 0|b_{a,\bm p}\sqrt{(2\pi)^32\omega_{\bm p}}.
\eea It follows that 
\bea 
\langle a,\bm p|\bm p',b\rangle=(2\pi)^3 2\omega_{\bm p}\gamma_{ab}\delta(\bm p-\bm p').
\eea 
The corresponding radiative field in the Cartesian frame is 
\bea 
\langle A_a(u,\Omega)|=\langle 0|A_A(u,\Omega)e^A_a(\Omega)=-\frac{i}{8\pi^2}\int_0^\infty d\omega e^{-i\omega u}\langle a,\bm p|.\label{outA}
\eea 
Note that one may define another conjugate state 
\bea 
\overline{\langle a,\bm p|}=\left(|\bm p,a\rangle\right)^\dagger=\langle 0|c_{a,\bm p}\sqrt{(2\pi)^3 2\omega_{\bm p}}=\langle b,\bm p|\left(C^{-1}\right)^b_{\ a}=\langle b,\bm p|\left(C^*\right)^b_{\ a}
\eea and 
\bea 
\overline{\langle A_a(u,\Omega)|}=\left(|A_a(u,\Omega)\rangle\right)^\dagger=\langle 0|A_A(u,\Omega) \left(e^A_a\right)^*=\langle 0|A_A(u,\Omega)
\left(C^*\right)^b_{\ a}e^A_b=\langle A_b(u,\Omega)|\left(C^*\right)^b_{\ a}.\nn\\
\eea For one-particle state, the completeness relation 
\bea 
1=\int\frac{d^3\bm p}{(2\pi)^32\omega_{\bm p}}|\bm p,a\rangle\gamma^{ab}\langle b,\bm p|
\eea is transformed to  
\bea 
1=i\int du d\Omega \gamma^{ab}\left( |A_a(u,\Omega)\rangle\langle\dot{A}_b(u,\Omega)|-|\dot{A}_a(u,\Omega)\rangle\langle{A}_b(u,\Omega)|\right) \label{complete}
\eea with $\dot{A}=\partial_uA$. Integrating by parts and ignoring the boundary term\footnote{There could be boundary terms at $\mathcal{I}^{\pm}_{\pm}$ associated with zero energy modes which have non-trivial contribution to the completeness relation. However, if we focus on the contribution of finite energy modes, we may ignore them. For the treatment of soft modes, one may consult \cite{Strominger:2017zoo} and references therein. It would be interesting to explore the contribution of the soft modes to  the completeness relation. }, the completeness relation becomes 
\bea 
1=2i\int du d\Omega \gamma^{ab}|A_a(u,\Omega)\rangle\langle\dot{A}_b(u,\Omega)|=-2i\int du d\Omega \gamma^{ab}|\dot{A}_a(u,\Omega)\rangle\langle{A}_b(u,\Omega)|.
\eea 

\subsection{Carrollian amplitude}
In this part, we will define the Carrollian amplitude for the vector field utilizing the 
 relations \eqref{Aapa} and \eqref{outA}. At first, 
 we introduce a symbol $\sigma$ to distinguish the incoming and outgoing states 
\bea 
\sigma=\left\{\begin{array}{cc}+&\text{outgoing},\\
-&\text{incoming}.\end{array}\right.
\eea Therefore, a state with null momentum $p$ can be written as 
\be 
p^\mu=\sigma \omega n^\mu.
\ee We can also use $\sigma$ to distinguish the operators inserted at $\mathcal{I}^-$ and $\mathcal{I}^+$
\bea 
A_A(u,\Omega,\sigma)=\left\{\begin{array}{cc}A_A(u,\Omega)&\text{for}\ \sigma=+,\\ A^{\text{P}}_A(u,\Omega)&\text{for}\  \sigma=-.\end{array}\right.
\eea The Carrollian amplitude is the correlator at the boundary 
\bea 
i\, \mathcal{C}_{a_1\cdots a_n}(u_1,\Omega_1,\sigma_1;\cdots;u_n,\Omega_n,\sigma_n)=\langle A_{a_1}(u_1,\Omega_1,\sigma_1)\cdots A_{a_n}(u_n,\Omega_n,\sigma_n)\rangle.
\eea Similar to the scalar Carrollian amplitude, we can extract the connected part 
\bea 
&&i\, \mathcal{C}_{a_1\cdots a_n}(u_1,\Omega_1,\sigma_1;\cdots;u_n,\Omega_n,\sigma_n)\nn\\&=&\text{sign}_n\left(\frac{1}{8\pi^2 i}\right)^n \prod_{j=1}^n \int_0^\infty d\omega_j e^{-i\sigma_j\omega_j u_j}(2\pi)^4\delta^{(4)}(\sum_{j=1}^n p_j)i\, \mathcal{M}_{a_1\cdots a_n}(p_1,p_2,\cdots,p_n)
\eea where $\mathcal{M}_{a_1\cdots a_n}(p_1,p_2,\cdots,p_n)$ is the $\mathcal{M}$ matrix with momentum $p_1,p_2,\cdots,p_n$ and polarization $\epsilon^\mu_{a_1},\epsilon^{\mu_2}_{a_2},\cdots,\epsilon^{\mu_n}_{a_n}$
\bea 
\mathcal{M}_{a_1\cdots a_n}(p_1,p_2,\cdots,p_n)=\mathcal{M}_{\mu_1\cdots\mu_n}(p_1,\cdots,p_n)\epsilon^{\mu_1}_{a_1}(\bm p_1)\cdots\epsilon^{\mu_n}_{a_n}(\bm p_n).
\eea The symbol $\text{sign}_n$ 
\be 
\text{sign}_n=\sigma_1\sigma_2\cdots\sigma_n
\ee coming from the phase choice of the vector field at the boundary.

\subsubsection{Poincar\'e invariance}
For spacetime translation
\be 
u'=u-c\cdot n,\quad \Omega'=\Omega,\quad A_a'(u',\Omega')=A_a(u,\Omega),
\ee we have the following transformation law for the Carrollian amplitude 
\bea 
\langle \prod_{j=1}^n A_{a_j}(u'_j,\Omega'_j,\sigma_j)\rangle=\langle \prod_{j=1}^n A_{a_j}(u_j,\Omega_j,\sigma_j)\rangle.
\eea This is proved by using the invariance of the $\mathcal{M}$ matrix under spacetime translation and the conservation of four-momentum. 

For the Lorentz transformation 
\bea 
u'=\Gamma^{-1}u,\quad \Omega'=\Omega'(\Omega),\quad A'_a(u',\Omega')=\Gamma R_a^{\ b}A_b(u,\Omega),\label{Poincare}
\eea we find the following transformation law for the Carrollian amplitude 
\bea 
\langle \prod_{j=1}^n A_{a_j}(u'_j,\Omega'_j,\sigma_j)\rangle=\left(\prod_{j=1}^n \Gamma_j\  R_{a_j}^{\ b_j}\right) \langle \prod_{j=1}^n A_{b_j}(u_j,\Omega_j,\sigma_j)\rangle.
\eea 
\paragraph{Proof.} 
\bea 
\text{LHS}&=&\text{sign}_n\left(\frac{1}{8\pi^2i}\right)^n\prod_{j=1}^n\int_0^\infty d\omega'_j e^{-i\sigma_j\omega'_j u'_j}(2\pi)^4\delta^{(4)}(\sum_{j=1}^n p'_j)i\mathcal{M}_{a_1\cdots a_n}(p'_1,p'_2,\cdots,p'_n)\nn\\&=&\text{sign}_n\left(\frac{1}{8\pi^2i}\right)^n\prod_{j=1}^n \int_0^\infty \Gamma_j d\omega_j e^{-i\sigma_j \omega_j u_j}(2\pi)^4\delta^{(4)}(\sum_{j=1}^n p_j)i \mathcal{M}_{a_1\cdots a_n}(p_1',p_2',\cdots,p_n')
\nn\\&=&\text{sign}_n\left(\frac{1}{8\pi^2i}\right)^n\prod_{j=1}^n \int_0^\infty \Gamma_j d\omega_j e^{-i\sigma_j \omega_j u_j}\Lambda_{\mu_j}^{\ \nu_j}\epsilon_{a_1}^{\mu_j}(p_j')(2\pi)^4\delta^{(4)}(\sum_{j=1}^n p_j)i \mathcal{M}_{\nu_1\cdots\nu_n}(p_1,p_2,\cdots,p_n).\nn\\ \label{lhs}
\eea In the second line, we have used the transformation law of the energy and retarded time
\be 
\omega_j'=\Gamma_j \omega_j,\quad u_j'=\Gamma_j^{-1}u_j
\ee and the invariance of the Dirac delta function 
\be 
\delta^{(4)}(\sum_{j=1}^n p_j')=\delta^{(4)}(\sum_{j=1}^n p_j)
\ee under Lorentz transformation. In the third line, we have used the covariance of the $\mathcal{M}$ matrix. Note that the polarization vector $\epsilon^\mu_a(\bm p)$ only depends on the angular part of the momenta, so we may rewrite it as 
\be 
\epsilon^\mu_a(\bm p)=\epsilon^\mu_a(\Omega).
\ee Since the Lorentz transformation leaves the metric $\gamma_{AB}$ invariant, we have 
\bea 
\epsilon^\mu_a(\Omega')&=&\epsilon'^{\mu}_a(\Omega')=-Y'^{\mu}_{A}(\Omega')e'^{A}_{a}(\Omega').
\eea At the last step, we have used the relation \eqref{veil} between the polarization vector and the vielbein field. Recalling the transformation law of the conformal Killing vector \eqref{Y} and the vielbein field \eqref{v}, we find 
\bea 
\Lambda_\mu^{\ \rho}\epsilon^\mu_a(\Omega')&=&-\Lambda_\mu^{\ \rho}\Gamma^{-1}\frac{\partial\theta^B}{\partial\theta'^A}\Lambda^{\mu}_{\ \nu}\left(Y^\nu_B(\Omega)+ n^\nu\nabla_B\log\Gamma\right)\Gamma \frac{\partial\theta'^A}{\partial\theta^C}R_a^{\ b}(\Omega)e^C_{\ b}(\Omega)\nn\\&=&R_a^{\ b}(\Omega)(\epsilon^\rho_b(\Omega)-n^\rho e^B_b(\Omega)\nabla_B\log\Gamma).
\eea Substituting the above equation into \eqref{lhs}, the terms containing the null vector $n^\mu$ vanishes due to the Ward-Takahashi identity 
\bea 
\mathcal{M}_{\mu_1\cdots\mu_n}(p_1,\cdots,p_n)n^{\mu_j}=0,\quad j=1,2,\cdots,n.
\eea Therefore, only one term is left
\begin{align}
    \langle \prod_{j=1}^n A_{a_j}(u'_j,\Omega'_j,\sigma_j)\rangle&=\text{sign}_n\left(\frac{1}{8\pi^2i}\right)^n\prod_{j=1}^n \int_0^\infty \Gamma_j d\omega_j e^{-i\sigma_j \omega_j u_j}R_{a_j}^{\ b_j}(2\pi)^4\delta^{(4)}(\sum_{j=1}^n p_j)i\, \mathcal{M}_{b_1\cdots b_n}(p_1,p_2,\cdots,p_n)\nn\\
&=\left(\prod_{j=1}^n \Gamma_j\, R_{a_j}^{\ b_j}\right)\langle \prod_{j=1}^n A_{b_j}(u_j,\Omega_j,\sigma_j)\rangle.\label{Car}
\end{align}

Compared with the transformation law of the scalar Carrollian amplitude 
\bea 
\langle \prod_{j=1}^n \Sigma(u_j',\Omega_j',\sigma_j)\rangle=\left(\prod_{j=1}^n \Gamma_j\right)\langle \prod_{j=1}^n \Sigma(u_j,\Omega_j,\sigma_j)\rangle,
\eea there is an extra local rotation associated with each vector field.  


\subsubsection{Propagators}
To compute the Carrollian amplitude in the position space, we need to the following propagators. The boundary to boundary propagator is just the two-point Carrollian amplitude\footnote{This is the electric Carrollian amplitude for the fundamental field. In the literature, there are discussions on the Carrollian CFTs which have conformal Carroll symmetries. Formally this is obtained by taking the $c\to 0$ limit of CFTs. In these Carrollian CFTs, the two-point Carrollian amplitudes are found by solving the corresponding Ward identities. There are electric and magnetic branches in the solutions \cite{deBoer:2021jej, Chen:2021xkw,Baiguera:2022lsw,Rivera-Betancour:2022lkc}. The electric branch is related to the radiative part and emerges naturally in the framework of bulk reduction. It would be interesting to understand the magnetic part which may be related to the soft part from bulk reduction.} which has been worked out in \cite{Liu:2023qtr}. We translate it into the following form 
\bea 
\langle A_a(u_1,\Omega_1,+)A_b(u_2,\Omega_2,-)\rangle=-\frac{1}{4\pi}I_0(\omega_0(u_1-u_2-i\epsilon))\gamma_{ab}\delta(\Omega_1-\Omega_2)
\eea where  
\be 
I_0(\omega_0 (u_1-u_2-i\epsilon))=\gamma_E+\log i\omega_0(u_1-u_2-i\epsilon)
\ee with $\gamma_E$ the Euler constant and $\omega_0>0$ is an IR cutoff. The external line from bulk $x$ to the boundary point at $(u,\Omega)$ at $\mathcal{I}^+$ is defined as\footnote{The name of the external line \cite{Liu:2024nfc} follows from standard textbook \cite{1995iqft.book.....P}. This is also called bulk-to-boundary propagator in the recent paper \cite{Alday:2024yyj} follows from the flat space limit of AdS propagator. We will use external line and bulk-to-boundary propagator interchangeably.}
\bea 
D_{\mu a}(u,\Omega,+;x)=\langle A_a(u,\Omega,+)a_\mu(x)\rangle=-\frac{\epsilon_{\mu a}(\Omega)}{8\pi^2(u+n\cdot x-i\epsilon)}
\eea where 
\be
\epsilon_{\mu a}=\gamma_{ab}\epsilon^b_\mu.
\ee Similarly, the external line from bulk to the boundary point $u,\Omega$ at $\mathcal{I}^-$ is 
\bea 
D_{\mu a}(u,\Omega,-;x)=\langle a_\mu(x)A_a(u,\Omega,-)\rangle=-\frac{\epsilon_{\mu a}(\Omega)}{8\pi^2(u+n\cdot x+i\epsilon)}.
\eea Therefore, the external line is unified as  
\bea 
D_{\mu a}(u,\Omega,\sigma;x)=-\frac{\epsilon_{\mu a}(\Omega)}{8\pi^2(u+n\cdot x-i\sigma\epsilon)}.
\eea 
The complex conjugate of the external line is 
\bea 
\left(D_{\mu a}(u,\Omega,\sigma;x)\right)^*=C_{a}^{\ c}D_{\mu c}(u,\Omega,-\sigma;x).
\eea 
There is a discontinuity 
\bea 
D_{\mu a}(u,\Omega,+;x)-D_{\mu a}(u,\Omega,-;x)=-\frac{i}{4\pi}\epsilon_{\mu a}(\Omega)\delta(u+n\cdot x).
\eea 
To obtain the bulk-to-boundary propagator, we may use the completeness relation 
\bea 
|a_\mu(x)\rangle=a_\mu(x)|0\rangle=\int du d\Omega \gamma^{ab}K_{\mu,a}(u,\Omega,-;x)|A_b(u,\Omega)\rangle
\eea from which we can read out the bulk-to-boundary propagator 
\bea 
K_{\mu a}(u,\Omega,-;x)=2i \partial_u D(u,\Omega,-;x)=\frac{i\epsilon_{\mu a}(\Omega)}{4\pi^2(u+n\cdot x+i\epsilon)^2}.
\eea For the conjugate state, we have 
\bea 
\langle a_\mu(x)|=\langle 0|a_\mu(x)=\int du d\Omega \gamma^{ab}K_{\mu a}(u,\Omega,+;x)\langle A_b(u,\Omega)|
\eea where 
\bea 
K_{\mu a}(u,\Omega,+;x)=-2i \partial_u D(u,\Omega,+;x)=-\frac{i\epsilon_{\mu a}(\Omega)}{4\pi^2(u+n\cdot x-i\epsilon)^2}.
\eea Therefore, 
\bea 
K_{\mu a}(u,\Omega,\sigma;x)=-\frac{i\sigma\epsilon_{\mu a}(\Omega)}{4\pi^2(u+n\cdot x-i\sigma\epsilon)^2}.
\eea

\subsubsection{MHV amplitude for four gluons} 
The MHV amplitude for four gluons has been studied in \cite{Mason:2023mti}. We will work on this example in our framework with the conventions reviewed in Appendix \ref{sh} and check the formalism presented in previous sections. The color-ordered MHV amplitude for four gluons is \cite{Parke:1986gb}
\bea 
\mathcal{M}[1^{-1},2^{-1},3^{+1},4^{+1}]=\frac{\langle 12\rangle^4}{\langle 12\rangle \langle 23\rangle \langle 34\rangle \langle 41\rangle}.\label{MHV4gluon}
\eea With the Lorentz transformation of the angle and square brackets, we find 
\bea 
\mathcal{M}[1'^{-1},2'^{-1},3'^{+1},4'^{+1}]=t_1^*t_2^*t_3 t_4 \mathcal{M}[1^{-1},2^{-1},3^{+1},4^{+1}].
\eea 
Note that  the gluon amplitude \eqref{MHV4gluon} corresponds to the amplitude with lower indices\footnote{With the flat metric $\gamma_{ab}$, one can find the polarization vector 
\be 
\epsilon^\mu_+=e^{2i\varphi}\frac{\langle q|\gamma_\mu|p]}{\sqrt{2}\langle qp\rangle},\quad \epsilon^\mu_-=-e^{-2i\varphi}\frac{[q|\gamma_\mu|p\rangle}{\sqrt{2}[qp]}\nn
\ee which 
 matches with the convention in \cite{2015sagt.book.....E} up to a phase.
}
\be 
\mathcal{M}_{-,-,+,+}[1,2,3,4]=\mathcal{M}[1^{-1},2^{-1},3^{+1},4^{+1}].
\ee 
Substituting into the definition of the Carrollian amplitude, we find 
\bea 
\mathcal{C}_{-,-,+,+}[u_1',\Omega_1',\sigma_1;\cdots;u_4',\Omega_4',\sigma_4]=\Gamma_1\Gamma_2\Gamma_3\Gamma_4 t_1^*t_2^*t_3t_4 \mathcal{C}_{-,-,+,+}[u_1,\Omega_1,\sigma_1;\cdots;u_4,\Omega_4,\sigma_4],
\eea which is exactly the transformation law \eqref{Car} with diagonal matrix $R_a^{\ b}$ \eqref{Rab}. 
To simplify notation, we will use the following abbreviation \be  \mathcal{C}_{a_1a_2a_3a_4}(1,2,3,4)=\mathcal{C}_{a_1a_2a_3a_4}(u_1,\Omega_1,\sigma_1;\cdots;u_4,\Omega_4,\sigma_4).\label{3135}
\ee When the amplitude is color ordered, we change the round brackets to square brackets in the notation 
\be 
\mathcal{C}_{a_1a_2a_3a_4}[1,2,3,4]=\mathcal{C}_{a_1a_2a_3a_4}[u_1,\Omega_1,\sigma_1;\cdots;u_4,\Omega_4,\sigma_4].
\ee 
For four point amplitude, we can fix $z_1=0,\ z_2=1,\ z_3=\infty,\ z_4=z$ by M\"obius transformation. In this case, we have\footnote{The cross ratio \be z=-\frac{z_{14}z_{23}}{z_{12}z_{34}}\nn\ee  is invariant under M\"obius transformation which is not the one used in \cite{Liu:2024nfc} where the cross ratio is defined as 
\be 
{Z}=\frac{z_{12}z_{34}}{z_{13}z_{24}}=\frac{1}{1-z}.\nn
\ee  }
\bs\begin{align}
    \langle 12\rangle&=-\sqrt{2\omega_1\omega_2},\\
    \langle 23\rangle&=-\sqrt{2\omega_2\omega_3},\\
    \langle 34\rangle&=\sqrt{\frac{4\omega_3\omega_4}{1+|z|^2}},\\
    \langle 41\rangle&=\sqrt{\frac{4\omega_1\omega_4}{1+|z|^2}}\bar z, 
\end{align}\es and then 
\bea 
\mathcal{M}_{-,-,+,+}[1,2,3,4]=\frac{\omega_1\omega_2}{\omega_3\omega_4}\frac{1+|z|^2}{2\bar z}.
\eea The conservation of the four-momentum leads to the Dirac delta function 
\be
\delta^{(4)}(\sum_{j=1}^4 p_j)=\frac{1+z^2}{2\omega_4}\delta(\omega_1-\omega_1^0)\delta(\omega_2-\omega_2^0)\delta(\omega_3-\omega_3^0)\delta(\bar z-z)
\ee with
\bs\begin{align}
   \omega_1^0&=-\sigma_1\sigma_4\frac{1-z}{1+z^2}\omega_4,\\
   \omega_2^0&=-\sigma_2\sigma_4\frac{2z}{1+z^2}\omega_4,\\
   \omega_3^0&=\sigma_3\sigma_4\frac{z(1-z)}{1+z^2}\omega_4.
\end{align}\es Therefore, 
\bea 
\mathcal{M}_{-,-,+,+}[1,2,3,4]=\sigma_1\sigma_2\sigma_3\sigma_4\frac{1}{z}.
\eea The Carrollian amplitude is 
\bea 
&&\mathcal{C}_{-,-,+,+}[1,2,3,4]\nn\\&=&\text{sign}_4\left(\frac{1}{8\pi^2i}\right)^4 \prod_{j=1}^4  \int_0^\infty d\omega_j e^{-i\sigma_j\omega_j u_j}(2\pi)^4\delta^{(4)}(\sum_{j=1}^4 p_j) \mathcal{M}_{-,-,+,+}[1,2,3,4]\nn\\&=&\frac{1}{(4\pi)^4}\Theta(\omega_1^0)\Theta(\omega_2^0)\Theta(\omega_3^0)\delta(\bar z-z)\frac{1+z^2}{2z}\int_0^\infty \frac{d\omega_4}{\omega_4}e^{-i\sigma_4\omega_4\chi},
\eea where 
\bea 
\chi=u_4-\frac{1-z}{1+z^2}u_1-\frac{2z}{1+z^2}u_2+\frac{z(1-z)}{1+z^2}u_3
\eea which is translation invariant. For $\sigma_1=\sigma_2=-,\sigma_3=\sigma_4=+$, the product of the step functions becomes 
\be 
\Theta(\omega_1^0)\Theta(\omega_2^0)\Theta(\omega_3^0)=\Theta(z)\Theta(1-z).
\ee The integration can be regularized by introducing an $IR $ cutoff $\omega_0$, then the Carrollian amplitude becomes 
\bea 
\mathcal{C}_{-,-,+,+}[1,2,3,4]=-\frac{1}{(4\pi)^4}\Theta(z)\Theta(1-z)\delta(\bar z-z)\frac{1+z^2}{2z}I_0(\omega_0\chi).
\eea To transform back to the general reference frame, we should use the M\"obius transformation which is parameterized by 
\bea 
a=\mp \sqrt{\frac{z_{21}}{z_{13}z_{23}}}z_3,\quad b=\pm \sqrt{\frac{z_{23}}{z_{21}z_{13}}}z_1,\quad c=\mp \sqrt{\frac{z_{21}}{z_{13}z_{23}}},\quad d=\pm \sqrt{\frac{z_{23}}{z_{21}z_{13}}}.
\eea The corresponding redshift factors are 
\bs\begin{align}
    \Gamma_1&=(1+|z_1|^2)|d|^2=(1+|z_1|^2)\Big|\frac{z_{23}}{z_{12}z_{13}}\Big|,\\
    \Gamma_2&=\frac{1+|z_2|^2}{2}|c+d|^2=\frac{1+|z_2|^2}{2}\Big|\frac{z_{13}}{z_{12}z_{23}}\Big|,\\
    \Gamma_3&=(1+|z_3|^2)|c|^2=(1+|z_3|^2)\Big|\frac{z_{12}}{z_{13}z_{23}}\Big|,\\
    \Gamma_4&=\frac{1+|z_4|^2}{1+|z|^2}\Big|c z+d\Big|^2.
\end{align}\es  The rotation factor $t_i$ can be obtained by \eqref{littlet}
\be 
    t_1=\frac{\bar d}{d},\quad
    t_2=\frac{\bar c+\bar d}{c+d},\quad
    t_3=\frac{\bar c}{c},\quad
    t_4=\frac{\bar c \bar z+\bar d}{cz+d}.
\ee Therefore, we find the following factor coming from Lorentz transformation with the help of \eqref{B27}
\bea \label{gammat}
\Gamma_1\Gamma_2\Gamma_3\Gamma_4t_1^*t_2^* t_3 t_4=\frac{1}{2(1+|z|^2)}\bar c^2 d^2 (c+d)^2(\bar c\bar z+\bar d)^2\prod_{j=1}^4 (1+|z_j|^2).
\eea

 We can also exchange the role of $1$ and $3$, and then get the color ordered amplitude
\bea 
\mathcal{M}_{+,-,-,+}[1,2,3,4]=\frac{\langle 23\rangle^4}{\langle 12\rangle \langle 23\rangle\langle 34\rangle\langle 41\rangle}.\label{pmmp}
\eea By fixing $z_1=0,\ z_2=1,\ z_3=\infty,\ z_4=z$, and utilizing the four-momentum conservation, we find 
\bea 
\mathcal{M}_{+,-,-,+}[1,2,3,4]=\sigma_1\sigma_2\sigma_3\sigma_4 z.
\eea Then the Carrollian amplitude is 
\bea 
\mathcal{C}_{+,-,-,+}[1,2,3,4]=-\frac{1}{(4\pi)^4}\Theta(z)\Theta(1-z)\delta(\bar z-z)\frac{z(1+z^2)}{2}I_0(\omega_0\chi).
\eea Transforming back to the general frame, 
\bea 
 &&\mathcal{C}_{+,-,-,+}[1',2',3',4']\nn\\&=&\Gamma_1\Gamma_2\Gamma_3\Gamma_4t_1t_2^*t_3^* t_4\mathcal{C}_{+,-,-,+}[1,2,3,4]\nn\\
   &=&-\frac{1}{(4\pi)^4}\Theta(z)\Theta(1-z)\delta(\bar z-z)\frac{z}{4}c^2 \bar d^2 (c+d)^2(\bar c\bar z+\bar d)^2\prod_{j=1}^4 (1+|z_j|^2)I_0(\omega_0\chi)\nn\\
  &=&-\frac{1}{(4\pi)^4}\Theta(z)\Theta(1-z)\delta(\bar z-z)\frac{z \bar z_{23}^{2}}{4 z_{23}^{2} \bar z_{34}^2 \bar z_{21}^{2}}\prod_{j=1}^4 (1+|z_j|^2)I_0(\omega_0\chi).
\eea In the last line, we have used the identity 
\be 
cz+d=\mp\sqrt{\frac{z_{23}z_{13}}{z_{21}z_{34}^2}}.
\ee 
Note that we should still change $u_i'\to u_i$ in the above equation to obtain 
\bea 
 &&\mathcal{C}_{+,-,-,+}[u_1,z_1,\bar{z}_1,\sigma_1;\cdots;u_4,z_4,\bar{z}_4,\sigma_4]\nn\\&=&-\frac{1}{(4\pi)^4}\Theta(z)\Theta(1-z)\delta(\bar z-z)\frac{z \bar z_{23}^{2}}{4 z_{23}^{2} \bar z_{34}^2 \bar z_{21}^{2}}\prod_{j=1}^4 (1+|z_j|^2)I_0(\tilde{\omega}_0\tilde{\chi})
\eea where we have used the Lorentz transformation law
\be 
\chi(\Gamma_ju_j)=\Gamma_4\tilde{\chi},\quad \tilde{\omega}_0=\Gamma_4^{-1}\omega_0
\ee with
\bea 
\tilde{\chi}=u_4-\Gamma_1\Gamma_4^{-1}\frac{1-z}{1+z^2}u_1-\Gamma_2\Gamma_4^{-1}\frac{2z}{1+z^2}u_2+\Gamma_3\Gamma_4^{-1}\frac{z(1-z)}{1+z^2}u_3.
\eea For upper indices, we find 
\bea 
&&\mathcal{C}^{+,-,-,+}[u_1,z_1,\bar{z}_1,\sigma_1;\cdots;u_4,z_4,\bar{z}_4,\sigma_4]\nn\\&=&-\frac{1}{(4\pi)^4}\Theta(z)\Theta(1-z)\delta(\bar z-z)\frac{z  z_{23}^{2}}{4 \bar z_{23}^{2}  z_{34}^2  z_{21}^{2}}\prod_{j=1}^4 (1+|z_j|^2)I_0(\tilde{\omega}_0\tilde{\chi}).\label{cpmmp}
\eea The comparison with \cite{Mason:2023mti} could be found in Appendix \ref{comparison}.

\section{Gravitational field}
The discussion for the vector field can be extended to the gravitational field straightforwardly. We work out the results in this section. 
\subsection{Setup}
For a symmetric transverse free gravitational field $h_{\mu\nu}$ 
with the fall-off condition 
\bea 
h_{\mu\nu}(x)=\frac{H_{\mu\nu}(u,\Omega)}{r}+\cdots
\eea near $\mathcal{I}^+$, one can extract the shear tensor 
\bea 
C_{AB}(u,\Omega)=Y^\mu_{A}Y^\nu_{B}H_{\mu\nu}(u,\Omega).\label{extractC}
\eea 
\paragraph{Poincar\'e transformation.} For the spacetime translation $x'^\mu=x^\mu+c^\mu$, the gravitational field is invariant 
\be 
h_{\mu\nu}'(x')=h_{\mu\nu}(x)
\ee which leads to 
\bea 
C_{AB}'(u',\Omega')=C_{AB}(u,\Omega),\quad u'=u-c\cdot n,\quad \Omega'=\Omega.
\eea Similarly, for the 
 Lorentz transformation 
\be 
h'_{\mu\nu}(x')=\Lambda_\mu^{\ \rho}\Lambda_\nu^{\ \sigma}h_{\rho\sigma}(x),
\ee we find 
\bea 
C'_{AB}(u',\Omega')=\Gamma^{-1}\frac{\partial\theta^C}{\partial\theta'^A}\frac{\partial\theta^D}{\partial\theta'^B}C_{CD}(u,\Omega),\quad C'^{AB}(u',\Omega')=\Gamma^3\frac{\partial\theta'^A}{\partial\theta^C}\frac{\partial\theta'^B}{\partial\theta^D}C^{CD}(u,\Omega).
\eea 
\textbf{Proof.} We will use \eqref{Y} and \eqref{extractC} as well as the relation $r'=\Gamma r$
\bea 
C'_{AB}(u',\Omega')&=&\Gamma^{-1}\frac{\partial\theta^C}{\partial\theta'^A}\left(Y^\mu_C(\Omega) + n^\mu\nabla_C\log\Gamma\right)\Gamma^{-1}\frac{\partial\theta^D}{\partial\theta'^B}\left(Y^\nu_D(\Omega) + n^\nu\nabla_D\log\Gamma\right)\Gamma H_{\mu\nu}(u,\Omega)\nn\\&=&\Gamma^{-1}\frac{\partial\theta^C}{\partial\theta'^A}\frac{\partial\theta^D}{\partial\theta'^B}C_{CD}(u,\Omega).
\eea In the second line, we have used the transverse condition for the gravitational field 
\be 
H_{\mu\nu}n^\nu=0.
\ee For the shear tensor with upper indices, we can use the transformation law \eqref{gammainverse} {to find}
\bea 
C'^{AB}(u',\Omega')=\gamma'^{AC}(\Omega')\gamma'^{BD}(\Omega')C'_{CD}(u',\Omega')=\Gamma^3\frac{\partial\theta'^A}{\partial\theta^C}\frac{\partial\theta'^B}{\partial\theta^D}C^{CD}(u,\Omega).
\eea 
Similar to the vector field, we can use the vielbein field to define the shear tensor in the local Cartesian frame 
\bea 
C_{ab}(u,\Omega)=e_a^A(\Omega)e_b^B(\Omega)C_{AB}(u,\Omega),\quad C^{ab}(u,\Omega)=e^a_A(\Omega)e^b_B(\Omega)C^{AB}(u,\Omega)\label{classC}
\eea 
whose transformation law under Lorentz transformation is 
\bea 
C_{ab}'(u',\Omega')=\Gamma R_a^{\ c}R_b^{\ d}C_{cd}(u,\Omega),\quad C'^{ab}(u',\Omega')=\Gamma R^a_{\ c}R^{b}_{\ d}C^{cd}(u,\Omega).
\eea 

\paragraph{Canonical quantization.} We will write the mode expansion as follows
\begin{align}
  h_{{\mu\nu}}(x)=&\int \frac{d^3\bm p}{\sqrt{(2\pi)^3}}\frac{1}{\sqrt{2\omega_{\bm p}}}[\epsilon^{ab}_{{\mu\nu}}(\bm p)b_{ab,\bm p}e^{-i\omega t+i\bm p\cdot\bm x}+\epsilon^{ab}_{{\mu\nu}}(\bm p)c^\dagger_{ab,\bm p}e^{i\omega t-i\bm p\cdot\bm x}],\label{hmnexp}
\end{align} where the polarization tensors should satisfy the symmetric and traceless conditions
\be 
\epsilon_{\mu\nu}^{ab}=\epsilon_{\mu\nu}^{ba}=\epsilon_{\nu\mu}^{ab},\quad \epsilon^{ab}_{\mu\nu}\gamma^{\mu\nu}=0,\quad \epsilon^{ab}_{\mu\nu}\gamma_{ab}=0
\ee and  the  orthogonality  and completeness relations  are 
\bea 
\epsilon_{\mu\nu}^{ab}\epsilon_{\rho\sigma}^{cd}\gamma^{\mu\nu,\rho\sigma}=\gamma^{ab,cd},\quad \epsilon_{\mu\nu}^{ab}\epsilon_{\rho\sigma}^{cd}\gamma_{ab,cd}=\gamma_{\mu\nu,\rho\sigma}
\eea where 
\bea 
\gamma_{ab,cd}=\frac{1}{2}(\gamma_{ac}\gamma_{bd}+\gamma_{ad}\gamma_{b c}-\gamma_{ab}\gamma_{cd}).
\eea Given the polarization vectors of the vector field which satisfy \eqref{orthocomp}, the polarization tensors of the gravitational field may be chosen as 
\bea 
\epsilon_{\mu\nu}^{ab}=\frac{1}{2}(\epsilon_\mu^a\epsilon_\nu^b+\epsilon_\mu^b\epsilon_\nu^a-\gamma_{\mu\nu}\gamma^{ab}).
\eea The annihilation and creation operators obey the commutation relations
\bea 
[b_{ab,\bm p},b_{cd,\bm p'}]=[c_{ab,\bm p}^\dagger, c_{cd,\bm p'}^\dagger]=0,\quad [b_{ab,\bm p},c_{cd,\bm p'}^\dagger]=\gamma_{ab,cd}\delta(\bm p-\bm p').
\eea 
For the gravitational field to be real, these operators should satisfy  the relations 
\bea 
c_{cd,\bm p}^\dagger=C^a_{\ c}C^b_{\ d}b_{ab,\bm p}^\dagger,\quad b_{cd,\bm p}=C^a_{\ c}C^b_{\ d}c_{ab,\bm p}.
\eea 
The mode expansion of the shear tensor is 
\bea 
C_{AB}(u,\Omega)&=&-i\int_0^\infty \frac{d\omega}{\sqrt{4\pi}}\sqrt{\omega}Y^\mu_A Y^\nu_B \epsilon^{ab}_{\mu\nu}(\Omega) (b_{ab,\bm p}e^{-i\omega u}- c^\dagger_{ab,\bm p}e^{i\omega u}).
\eea We may write 
\be 
e_{AB}^{ab}=Y^\mu_A Y^\nu_B \epsilon_{\mu\nu}^{ab}=\frac{1}{2}(e_A^a e_B^b+e_A^b e_B^a-\gamma_{AB}\gamma^{ab})\label{vielbeinab}
\ee  and then 
\bea 
|C_{AB}(u,\Omega)\rangle=\frac{i}{8\pi^2}\int_0^\infty d\omega e^{i\omega u} e_{AB}^{ab}|\bm p,ab\rangle,
\eea where we have defined the state 
\bea 
|\bm p,ab\rangle=\sqrt{(2\pi)^3 2\omega_{\bm p}}c_{ab,\bm p}^\dagger|0\rangle.
\eea The inverse of $e_{AB}^{ab}$ is 
\be 
e^{AB}_{ab}=\frac{1}{2}(e^A_{a}e^B_b+e^A_be^B_a-\gamma^{AB}\gamma_{ab}),
\ee thus we can define a state 
\bea 
|C_{ab}(u,\Omega,+)\rangle=e^{AB}_{ab}|C_{AB}(u,\Omega)\rangle=\frac{i}{8\pi^2}\int_0^\infty d\omega  e^{i\omega u}|\bm p,ab\rangle.
\eea 
Since the shear tensor is symmetric and traceless, we can also write the state $|C_{ab}(u,\Omega)\rangle$ as 
\bea 
|C_{ab}(u,\Omega,+)\rangle=e^A_a e^B_b|C_{AB}(u,\Omega)\rangle
\eea which is consistent with the classical definition of $C_{ab}$ in \eqref{classC}. We can also find the following state 
\bea 
\langle C_{ab}(u,\Omega)|=-\frac{i}{8\pi^2}\int_0^\infty d\omega e^{-i\omega u}\langle ab,\bm p|
\eea with 
\bea 
\langle ab,\bm p|=\langle 0|b_{ab,\bm p}\sqrt{(2\pi)^3 2\omega_{\bm p}}.
\eea 
 Near $\mathcal{I}^-$, we may define the state by antipodal map 
 \bea 
 |C_{AB}(u,\Omega,-)\rangle=|C_{AB}^{\text{P}}(u,\Omega)\rangle=-\frac{i}{8\pi^2}\int_0^\infty d\omega e^{i\omega u}e^{ab}_{AB}(\Omega)|\bm p;ab\rangle
 \eea 
 where 
 \bea 
 C_{AB}^{\text{P}}(u,\Omega)=\left(Y^{\text{P}}\right)^\mu_A(\Omega)\left(Y^{\text{P}}\right)_B^\nu(\Omega)H_{\mu\nu}^{\text{P}}(u,\Omega)=Y^\mu_A(\Omega)Y^\nu_B(\Omega)H_{\mu\nu}(v\to u,\Omega^{\text{P}}).
 \eea 
 
    Therefore, 
    \bea 
|C_{ab}(u,\Omega,-)\rangle=e^{AB}_{ab}(\Omega)|C_{AB}(u,\Omega,-)\rangle=e^A_ae^B_b|C_{AB}(u,\Omega,-)\rangle=-\frac{i}{8\pi^2}\int_0^\infty d\omega e^{i\omega u}|\bm p;ab\rangle.
    \eea 
    
    \paragraph{Carrollian amplitude.}
    The Carrollian amplitude is 
    \bea 
    &&i\, \mathcal{C}_{a_1b_1,a_2b_2,\cdots,a_nb_n}(u_1,\Omega_1,\sigma_1;\cdots;u_n,\Omega_n,\sigma_n)=\langle\prod_{j=1}^n C_{a_jb_j}(u_j,\Omega_j,\sigma_j)\rangle\nn\\&=&\left(\frac{1}{8\pi^2i}\right)^n \prod_{j=1}^n \int_0^\infty d\omega_j e^{-i\sigma_j\omega_j u_j}(2\pi)^4 \delta^{(4)}(\sum_{j=1}^n p_j)i\, \mathcal{M}_{a_1b_1,a_2b_2,\cdots,a_nb_n}(p_1,p_2,\cdots,p_n).
    \eea At the second step, we just write out the $\mathcal{M}$ matrix, ignoring the amplitude which is from the identity operator. It is easy to prove the transformation law of the Carrollian amplitude under spacetime translation 
   \bea 
   \mathcal{C}_{a_1b_1,\cdots,a_nb_n}'(u_1',\Omega_1',\sigma_1;\cdots;u_n',\Omega_n',\sigma_n)=\mathcal{C}_{a_1b_1,\cdots,a_nb_n}(u_1,\Omega_1,\sigma_1;\cdots;u_n,\Omega_n,\sigma_n).
   \eea For the Lorentz transformation, utilizing the Ward-Takahashi identity as the derivation of \eqref{Car}, one can find 
   \bea 
   \mathcal{C}_{a_1b_1,\cdots,a_nb_n}'(u_1',\Omega_1',\sigma_1;\cdots;u_n',\Omega_n',\sigma_n)=\left(\prod_{j=1}^n \Gamma_j R_{a_j}^{\ c_j}R_{b_j}^{\ d_j}\right)\mathcal{C}_{c_1d_1,\cdots,c_nd_n}(u_1,\Omega_1,\sigma_1;\cdots;u_n,\Omega_n,\sigma_n).\nn\\\label{Carrgra}
   \eea 
   \paragraph{Progapators.}
   The boundary-to-boundary propagator is the two-point Carrollian amplitude 
   \bea 
   \langle C_{ab}(u_1,\Omega_1,+)C_{cd}(u_2,\Omega_2,-)\rangle=-\frac{1}{4\pi}I_0(\omega_0(u_1-u_2-i\epsilon))\gamma_{ab,cd}\delta(\Omega_1-\Omega_2).
   \eea The external line from bulk $x$ to the boundary point at $(u,\Omega)$ at $\mathcal{I}^+$ is 
\bea 
D_{\mu\nu, ab}(u,\Omega,+;x)=\langle C_{ab}(u,\Omega,+)h_{\mu\nu}(x)\rangle=-\frac{\epsilon_{\mu\nu ab}(\Omega)}{8\pi^2(u+n\cdot x-i\epsilon)}\label{Dp}
\eea and the external line from bulk point $x$ to the boundary point at $\mathcal{I}^-$ is 
\be 
D_{\mu\nu,ab}(u,\Omega,-;x)=\langle h_{\mu\nu}(x)C_{ab}(u,\Omega,-)\rangle=\frac{\epsilon_{\mu\nu ab}(\Omega)}{8\pi^2(u+n\cdot x+i\epsilon)}.\label{dm}
\ee The two external lines \eqref{Dp} and \eqref{dm} are unified to 
\bea 
D_{\mu\nu,ab}(u,\Omega,\sigma;x)=-\frac{\sigma \epsilon_{\mu\nu ab}(\Omega)}{8\pi^2(u+n\cdot x-i\sigma \epsilon)}.
\eea Under complex conjugate, we find 
\bea 
D^*_{\mu\nu,ab}(u,\Omega,\sigma;x)=-C_a^{\  c}C_b^{\ d}D_{\mu\nu,cd}(u,\Omega,-\sigma;x).
\eea 
\subsection{MHV amplitude for four gravitons}
In spinor helicity formalism, the tree-level scattering amplitude of four gravitons is 
\bea 
&&\mathcal{M}(1^{-2},2^{-2},3^{+2},4^{+2})=\frac{\langle 12\rangle^7 [12]}{\langle 13\rangle \langle 14\rangle \langle 23\rangle \langle 24\rangle \langle 34\rangle^2}=\frac{\langle 12\rangle^4[34]^4}{stu}=\frac{\langle 12\rangle^4[34]^4}{\langle 12\rangle [12]\langle 13\rangle [13]\langle 14\rangle [14]}.\nn\\
\eea As the four gluon scattering, we transform to the notation in this work 
\bea 
\mathcal{M}_{--,--,++,++}(1,2,3,4)=\frac{\langle 12\rangle^4[34]^4}{\langle 12\rangle [12]\langle 13\rangle [13]\langle 14\rangle [14]}
\eea and the Carrollian amplitude should transform in the Lorentz transformation as \eqref{Carrgra}.
It is enough to present the result in the reference frame with $z_1=0,\ z_2=1,\ z_3=\infty,\ z_4=z$ in which
\bs\begin{align}
    [12]&=-\langle 12\rangle=\sqrt{2\omega_1\omega_2},\\
    [13]&=-\langle 13\rangle=\sqrt{4\omega_1\omega_3},\\
    [14]&=\sqrt{\frac{4\omega_1\omega_4}{1+|z|^2}}z,\quad \langle 14\rangle=-\sqrt{\frac{4\omega_1\omega_4}{1+|z|^2}}\bar z,\\
    [34]&=-\langle 34\rangle=-\sqrt{\frac{4\omega_3\omega_4}{1+|z|^2}}.
\end{align}
\es Therefore, by imposing the four-momentum conservation
\bea 
\mathcal{M}_{--,--,++,++}(1,2,3,4)= -2\frac{\omega_2\omega_3\omega_4}{\omega_1}\frac{1}{(1+z^2)z^2}=-4\sigma_1\sigma_2\sigma_3\sigma_4 \frac{\omega_4^2}{(1+z^2)^2}.
\eea 
We will set $\sigma_1=\sigma_2=-,\sigma_3=\sigma_4=+$.  Similar  to the abbreviation in the gluon Carrollian amplitude \eqref{3135}, we will use the abbreviation \bea 
\mathcal{C}_{++,--,--,++}(1,2,3,4)=\mathcal{C}_{++,--,--,++}(u_1,z_1,\bar z_1,\sigma_1;\cdots;u_4,z_4,\bar{z}_4,\sigma_4).
\eea The Carrollian amplitude is 
\bea 
\mathcal{C}_{--,--,++,++}(1,2,3,4)&=&\frac{1}{(4\pi)^4}\Theta(z)\Theta(1-z)\delta(\bar z-z)\frac{1+z^2}{2}\int_0^\infty d\omega_4 \omega_4e^{-i\omega_4\chi}\times \frac{-4}{(1+z^2)^2}\nn\\&=&\frac{1}{(4\pi)^4}\Theta(z)\Theta(1-z)\delta(\bar z-z)\frac{2}{1+z^2}\frac{1}{\chi^2}.
\eea 

Exchanging the role of $1$ and $3$, we find the following amplitude 
\bea 
\mathcal{M}_{++,--,--,++}(1,2,3,4)&=&\frac{\langle 23\rangle^4 [14]^4}{stu}=\frac{\langle 23\rangle^4 [14]^4}{\langle 12\rangle [12]\langle 13\rangle [13]\langle 14\rangle [14]}.
\eea 
Substituting the four-momentum conservation, we find 
\bea 
\mathcal{M}_{++,--,--,++}(1,2,3,4)=-\frac{2z^2}{1+z^2}\frac{\omega_2\omega_3\omega_4}{\omega_1}=-\sigma_1\sigma_2\sigma_3\sigma_4\frac{4z^4}{(1+z^2)^2}\omega_4^2.
\eea The Carrollian amplitude is 
\bea 
\mathcal{C}_{++,--,--,++}(1,2,3,4)=\frac{1}{(4\pi)^4}\Theta(z)\Theta(1-z)\delta(\bar z-z)\frac{2z^4}{1+z^2}\frac{1}{\chi^2}.
\eea Using Lorentz transformation, we find the Carrollian amplitude for general $z_1,z_2,z_3,z_4$
\bea 
&&\mathcal{C}_{++,--,--,++}(1',2',3',4')=\Gamma_1\Gamma_2\Gamma_3\Gamma_4 (t_1 t_2^* t_3^*t_4)^2\mathcal{C}_{++,--,--,++}(1,2,3,4)\nn\\&=&\frac{1}{(4\pi)^4}\Theta(z)\Theta(1-z)\delta(\bar z-z)\frac{c^3\bar d^3(c+d)^3(\bar c\bar z+\bar d)^3}{\bar c d (\bar c+\bar d)(cz+d)}\left(\prod_{j=1}^4 (1+|z_j|^2)\right) \frac{z^4}{(1+z^2)^2}\frac{1}{\chi^2}\nn\\&=&\frac{1}{(4\pi)^4}\Theta(z)\Theta(1-z)\delta(\bar z-z)\left(\prod_{j=1}^4 (1+|z_j|^2)\right)\frac{z_{12}z_{34}\bar{z}_{23}^4}{\bar z^3_{12}\bar z^3_{34} z_{23}^4} \frac{z^4}{(1+z^2)^2}\frac{1}{\chi^2}.
\eea 
To transform to the retarded time $u_j$ without prime, we should change $\chi=\chi(u_j)$ to $\chi(\Gamma_j u_j)$. Therefore, 
\bea 
&&\mathcal{C}_{++,--,--,++}(u_1,z_1,\bar{z}_1,\sigma_1;\cdots;u_4,z_4,\bar{z}_4,\sigma_4)\nn\\&=&\frac{1}{(4\pi)^4}\Theta(z)\Theta(1-z)\delta(\bar z-z)\left(\prod_{j=1}^4 (1+|z_j|^2)\right)\frac{z_{12}z_{34}\bar{z}_{23}^4}{\bar z^3_{12}\bar z^3_{34} z_{23}^4} \frac{z^4}{(1+z^2)^2}\frac{1}{\chi^2(\Gamma_j u_j)}\nn\\&=&\frac{1}{(4\pi)^4}\Theta(z)\Theta(1-z)\delta(\bar z-z)\left(\prod_{j=1}^4 (1+|z_j|^2)\right)\frac{z_{12}z_{34}\bar{z}_{23}^4}{\bar z^3_{12}\bar z^3_{34} z_{23}^4} \frac{z^4}{(1+z^2)^2}\frac{1}{\Gamma_4^2\tilde{\chi}^2}.
\eea 
For upper indices, we find 
\bea 
&&\mathcal{C}^{++,--,--,++}(u_1,z_1,\bar{z}_1,\sigma_1;\cdots,u_4,z_4,\bar{z}_4,\sigma_4)\nn\\&=&\frac{1}{(4\pi)^4}\Theta(z)\Theta(1-z)\delta(\bar z-z)\left(\prod_{j=1}^4 (1+|z_j|^2)\right)\frac{\bar z_{12}\bar z_{34}{z}_{23}^4}{ z^3_{12}z^3_{34} \bar z_{23}^4} \frac{z^4}{(1+z^2)^2}\frac{1}{\Gamma_4^2\tilde{\chi}^2}\nn\\&=&\frac{1}{(4\pi)^4}\Theta(z)\Theta(1-z)\delta(\bar z-z)z^4\frac{(1+|z_1|^2)(1+|z_2|^2)(1+|z_3|^2)}{1+|z_4|^2}\frac{z_{23}^3\bar{z}_{12}^2\bar{z}_{34}^3}{\bar{z}_{23}^5 z_{12}^2 z_{34}z_{13}\bar{z}_{13}}\frac{1}{\tilde{\chi}^2}.\label{fourgraviton}
\eea

\section{Higher dimensions}\label{hd}
In this section, we will extend the definition of the Carrollian amplitude to higher dimensions. In this case, the bulk Poincar\'e transformation remains the same form of four dimensions. The main difference is the fall-off conditions for the bulk field near $\mathcal{I}^{\pm}$ and that the boundary manifold becomes the $d-1$ dimensional Carrollian manifold $\mathbb{R}\times S^{d-2}$. One can find more details on the Carrollian manifold $\mathcal{I}^+$ in higher dimensions in our previous papers \cite{Li:2023xrr,Liu:2024rvz}.

\subsection{Scalar field}
For a bulk scalar field $\Phi(x)$ in $d$ dimensions, the fall-off behaviour may be \cite{Satishchandran:2019pyc}
\be 
\Phi(x)=\frac{\Sigma(u,\Omega)}{r^{\Delta}}+\cdots,
\ee where 
\be 
\Delta=\frac{d-2}{2}.
\ee 
For spacetime translation, the boundary field $\Sigma(u,\Omega)$ transforms as 
\be 
\Sigma'(u',\Omega')=\Sigma(u,\Omega),\quad u'=u-c\cdot n,\quad \bm n'=\bm n.\label{stSigmad}
\ee For Lorentz transformation, the boundary field $\Sigma(u,\Omega)$ is rescaled by 
\bea 
\Sigma'(u',\Omega')=\Gamma^\Delta \Sigma(u,\Omega),\quad u'=\Gamma^{-1}u,\quad \bm n'=\Gamma^{-1}\bm\Gamma.\label{LTSigmad}
\eea 
\paragraph{Infinitesimal transformation.} A general Carrollian diffeomorphism is generated by the vector 
\be 
\bm\xi_{f, Z}=f(u,\Omega)\partial_u+Z^A(\Omega)\partial_A
\ee and the transformation of the scalar field under Carrollian diffeomorphism is 
\bea 
\delta_{f,Z}\Sigma(u,\Omega)=f(u,\Omega)\partial_u\Sigma+Z^A(\Omega)\nabla_A\Sigma(u,\Omega)+\frac{1}{2}\nabla_CZ^C(\Omega)\Sigma(u,\Omega).
\eea 
We consider an infinitesimal spacetime translation at the boundary
\be 
u'=u-\epsilon\cdot n, \quad \bm n'=\bm n.
\ee The infinitesimal transformation of the field $\Sigma(u,\Omega)$ is 
\be 
\delta_{\epsilon}\Sigma(u,\Omega)=\Sigma'(u,\Omega)-\Sigma(u,\Omega)\approx \Sigma(u+\epsilon\cdot n,\Omega)-\Sigma(u,\Omega)\approx \epsilon\cdot n\dot\Sigma(u,\Omega).
\ee Therefore, a spacetime translation parameterized by $\epsilon^\mu$ is identified as the Carrollian diffeomorphism with
\be 
f=\epsilon \cdot n,\quad Z^A=0.
\ee 
Similarly, we can consider an infinitesimal Lorentz transformation at the boundary 
\bea 
u'=(1+\epsilon K)^{-1}u,\quad \theta'^A=\theta^A+\epsilon Y^A
\eea where the redshift factor is expanded around the identity
\be 
\Gamma\approx 1+\epsilon K.
\ee 
Since the redshift factor is generated by the conformal transformation, we should have 
\be 
K=-\frac{1}{d-2}\nabla_CY^C.
\ee 
The infinitesimal transformation of the field $\Sigma(u,\Omega)$ is 
\bea 
\delta_{\epsilon}\Sigma&=&\Sigma'(u,\Omega)-\Sigma(u,\Omega)\nn\\&\approx &\Gamma^\Delta \Sigma(\Gamma u,\Omega-\epsilon Y)-\Sigma(u,\Omega)\nn\\&\approx& (1+\epsilon K)^\Delta \Sigma(u(1+\epsilon K),\Omega-\epsilon Y)-\Sigma(u,\Omega)\nn\\&\approx&\epsilon u K \partial_u\Sigma-\epsilon Y^A\nabla_A\Sigma+ \epsilon\Delta K \Sigma(u,\Omega)\nn\\&=&- \epsilon\frac{u}{d-2}\nabla_CY^C\partial_u\Sigma-\epsilon (Y^A\nabla_A+\frac{1}{2}\nabla_CY^C)\Sigma.
\eea Therefore, the Lorentz transformation generated by a conformal Killing vector $Y^A$ is identified as the Carrollian diffeomorphism with
\be 
f=-\frac{u}{d-2}\nabla_CY^C,\quad Z^A=-Y^A.\label{CDLC}
\ee 
\paragraph{Quantization}
The mode expansion of the scalar field is \bea 
\Phi(x)=\int\frac{d^{d-1}\bm p}{\sqrt{(2\pi)^{d-1}}}\frac{1}{\sqrt{2\omega_{\bm p}}}(e^{-i\omega t+i\bm p\cdot\bm x}b_{\bm p}+e^{i\omega t-i\bm p\cdot\bm x}b_{\bm p}^\dagger)
\eea from which we may read out the mode expansion of the boundary field $\Sigma(u,\Omega)$ \cite{Li:2023xrr}
\bea 
\Sigma(u,\Omega)&=&\int_0^\infty \frac{d\omega}{\sqrt{4\pi\omega}}\omega^{\Delta}[e^{-i\pi \Delta/2}b_{\bm p}e^{-i\omega u}+e^{i\pi \Delta/2}b_{\bm p}^\dagger e^{i\omega u}]
\eea  and the boundary state 
\bea 
|\Sigma(u,\Omega,+)\rangle=\Sigma(u,\Omega)|0\rangle=e^{i\pi \Delta/2}\int_0^\infty \frac{d\omega}{\sqrt{(2\pi)^{d-1}\times 8\pi}}\omega^{\Delta-1}e^{i\omega u}|\bm p\rangle
\eea where the  state $|\bm p\rangle$ with definite momentum is 
\be 
|\bm p\rangle=\sqrt{(2\pi)^{d-1}2\omega_{\bm p}}\ b_{\bm p}^\dagger |0\rangle.
\ee 
The $+$ symbol is to denote the field at $\mathcal{I}^+$.
Similarly, we could find the boundary field at $\mathcal{I}^-$
\bea 
\Sigma(v,\Omega)=\int_0^\infty \frac{d\omega}{\sqrt{4\pi\omega}}\omega^{\Delta}[e^{i\pi \Delta/2}b_{\bm p^{\text{P}}}e^{-i\omega v}+e^{-i\pi \Delta/2}b^\dagger_{\bm p^{\text{P}}}e^{i\omega v}].
\eea By imposing the antipodal map, we find 
\bea 
\Sigma^{\text{P}}(u,\Omega)=\Sigma(v\to u,\Omega^{\text{P}})=\int_0^\infty \frac{d\omega}{\sqrt{4\pi\omega}}\omega^{\Delta}[e^{i\pi \Delta/2}b_{\bm p}e^{-i\omega u}+e^{-i\pi \Delta/2}b^\dagger_{\bm p}e^{i\omega u}].
\eea Therefore, we can define a state 
\bea 
|\Sigma(u,\Omega,-)\rangle=\Sigma^{\text{P}}(u,\Omega)|0\rangle=e^{-i\pi\Delta/2}\int_0^\infty \frac{d\omega}{\sqrt{(2\pi)^{d-1}\times 8\pi}}\omega^{\Delta-1}e^{i\omega u}|\bm p\rangle.
\eea 
Now  the scalar Carrollian amplitude in higher dimensions is 
\bea 
&&i\, \mathcal{C}(u_1,\Omega_1,\sigma_1;\cdots;u_n,\Omega_n,\sigma_n)=\langle \prod_{j=1}^n \Sigma(u_j,\Omega_j,\sigma_j)\rangle\nn\\&=&\left(\frac{(-1)^{\Delta}}{(2\pi)^{d-1}\times 8\pi}\right)^{n/2}\prod_{j=1}^n \int_0^\infty d\omega_j \omega_j^{\Delta-1}e^{-i\sigma_j\omega_j u_j}(2\pi)^{d}\delta^{(d)}(\sum_j p_j)i\mathcal{M}(p_1,p_2,\cdots, p_n).\label{Carrd}
\eea In the second line, we only extract the $\mathcal{M}$ matrix. As a consistency check, the scalar Carrollian amplitude reduces to the one in four dimensions for $d=4$. The main result is that the Carrollian amplitude is not the Fourier transform  of the scattering amplitude in general dimensions. The factor $\omega^{\Delta-1}$ is necessary to balance the dimensions in the definition. In $d$ dimensions, the $\mathcal{M}$ matrix has the  following mass dimension 
\be 
[\mathcal{M}]=d-n\Delta.
\ee As a consequence, the mass dimension on the right-hand side of \eqref{Carrd} is 
\be 
d-n\Delta-d+n\Delta=0
\ee which matches with the dimension of the Carrollian amplitude since the Carrollian amplitude is just the  probability amplitude in position space.

As in four dimensions, we can prove that 
\bea 
\langle \prod_{j=1}^n \Sigma(u_j',\Omega_j',\sigma_j)\rangle=\langle \prod_{j=1}^n \Sigma(u_j,\Omega_j,\sigma_j)\rangle\label{stC}
\eea for spacetime translation and 
\bea 
\langle \prod_{j=1}^n \Sigma(u_j',\Omega_j',\sigma_j)\rangle=\left(\prod_{j=1}^n \Gamma_j^\Delta\right)\langle \prod_{j=1}^n \Sigma(u_j,\Omega_j,\sigma_j)\rangle\label{LC}
\eea for Lorentz transformation.  The corresponding Ward identities could  be found in Appendix \ref{Ward}. Note that the formula \eqref{Carrd} is similar to the modified Mellin transform \cite{Banerjee:2018gce,Banerjee:2019prz,Banerjee:2024hvb} which is defined in four dimensions. However, the integral transform \eqref{Carrd} is defined in general dimensions.

\paragraph{Propagators.} The boundary-to-boundary propagator is the two-point Carrollian amplitude 
\bea 
\langle \Sigma(u_1,\Omega_1,+)\Sigma(u_2,\Omega_2,-)\rangle=\frac{e^{-i\pi\Delta}}{4\pi}I_0(\omega_0(u_1-u_2-i\epsilon))\delta(\Omega_1-\Omega_2).
\eea 
The result satisfies the transformation law under spacetime translation \eqref{stC}. To check the transformation law under Lorentz transformation \eqref{LC}, we can use \eqref{gamma} to find 
\bea 
\det \gamma'(\Omega')=\Gamma^{-2(d-2)}J^{-2}\det\gamma(\Omega),
\eea where $J$ is the Jacobian of the coordinate transformation  $\Omega\to\Omega'$
\be 
J=\det\frac{\partial\theta'^A}{\partial\theta^B}.
\ee Therefore, { we have}
\bea 
\delta(\Omega_1'-\Omega_2')=\frac{1}{\sqrt{\det\gamma'(\Omega')}}\delta(\bm\theta_1'-\bm\theta_2')=\frac{1}{\sqrt{\det\gamma'(\Omega')}J}\delta(\bm\theta_1-\bm\theta_2)=\Gamma^{2\Delta}\delta(\Omega_1-\Omega_2),
\eea and the two-point Carrollian amplitude satisfies the  transformation law \eqref{LC}.
Note that the phase factor $e^{-i\pi\Delta}$ may be absorbed into the definition of the boundary field. 

There are two kinds of bulk-to-boundary propagators 
\bs\begin{align}
    D(u,\Omega,+;x)&=\langle \Sigma(u,\Omega,+)\Phi(x)\rangle=\frac{e^{-i\pi\Delta} \Gamma(\Delta)}{4\pi\times (2\pi)^\Delta}\frac{1}{(u+n\cdot x-i\epsilon)^\Delta},\\
    D(u,\Omega,-;x)&=\langle \Phi(x)\Sigma(u,\Omega,-)\rangle=\frac{\Gamma(\Delta)}{4\pi\times(2\pi)^\Delta}\frac{1}{(u+n\cdot x+i\epsilon)^\Delta}.
\end{align}\es They satisfy the relation 
\bea 
D(u,\Omega,+;x)^*=e^{i\pi\Delta}D(u,\Omega,-;x).
\eea 

\paragraph{Completeness relation.}
 The completeness relation \bea 
1=\int\frac{d^{d-1}\bm p}{(2\pi)^{d-1}2\omega_{\bm p}}|\bm p\rangle\langle\bm p|
\eea is transformed to 
\bea 
1=i\int du d\Omega\left( |\Sigma(u,\Omega)\rangle\langle\dot{\Sigma}(u,\Omega)|-|\dot{\Sigma}(u,\Omega)\rangle\langle{\Sigma}(u,\Omega)|\right).
\eea Integrating by parts and ignoring the boundary term, the completeness relation becomes 
\bea 
1=2i\int du d\Omega |\Sigma(u,\Omega)\rangle\langle\dot{\Sigma}(u,\Omega)|=-2i\int du d\Omega |\dot{\Sigma}(u,\Omega)\rangle\langle{\Sigma}(u,\Omega)|.
\eea The result is exactly the same as the one in four dimensions.

\subsection{Spinning field}
We will mainly discuss the vector field in this part since it can be extended to general spinning theory straightforwardly. The vector field in $d$ dimensions has the fall-off behaviour near $\mathcal{I}^+$
\be 
a_{\mu}(x)=\frac{A_\mu(u,\Omega)}{r^\Delta}+\cdots.
\ee This condition should be supplemented by the constraint 
\be 
n^\mu A_\mu(u,\Omega)=0.
\ee  After some efforts, we can find that  the boundary field  $A_A(u,\Omega)$ transforms as 
\be 
A_A'(u',\Omega')=A_A(u,\Omega),\quad u'=u-c\cdot n,\quad \bm n'=\bm n
\ee for spacetime translations and 
\bea 
A_A'(u',\Omega')=\Gamma^{\Delta-1}\frac{\partial\theta^B}{\partial\theta'^A}A_B(u,\Omega),\quad A'^A(u',\Omega')=\Gamma^{\Delta+1}\frac{\partial\theta'^A}{\partial\theta^B}A^B(u,\Omega),\quad  u'=\Gamma^{-1}u,\quad \bm n'=\Gamma^{-1}\bm\Gamma\label{ALC}\nn\\
\eea  for Lorentz transformations. By introducing a vielbein field $e_A^a$ on $S^{d-2}$, we can define the vector field in the local Cartesian frame 
\be 
A_a(u,\Omega)=e_a^A(\Omega)A_A(u,\Omega),\quad A^a(u,\Omega)=e^a_A(\Omega)A^A(u,\Omega)
\ee that transforms as
\bea 
A'_a(u',\Omega')=\Gamma^\Delta R_a^{\ b}A_b(u,\Omega),\quad A'^a(u',\Omega')=\Gamma^\Delta R^a_{\ b}A^b(u,\Omega).\label{transALC}
\eea 

\paragraph{Infinitesimal transformation.} We can also identify the Poincar\'e transformation as a boundary Carrollian diffeomorphism. The infinitesimal variation of the field $A_A(u,\Omega)$ under Carrollian diffeomorphism is 
\bea 
\delta_{f,Z}A_A(u,\Omega)=f(u,\Omega)\partial_u A_A+Z^C\nabla_CA_A+\frac{1}{2}\nabla_CZ^CA_A+\frac{1}{2}(\nabla_AZ_C-\nabla_CZ_A)A^C.\label{CDA}
\eea We will only consider the Lorentz transformation since the spacetime translation is the same as the scalar theory. Utilizing  \eqref{ALC}, we find 
\bea 
\delta_\epsilon A_A&=&-\epsilon u \frac{1}{d-2}\nabla_CY^C\partial_u A_A-\epsilon (Y^C\nabla_CA_A+\nabla_AY^C A_C+\frac{\Delta-1}{2\Delta}\nabla_CY^C A_A)\nn\\&=&-\epsilon{u} \frac{1}{d-2}\nabla_CY^C\partial_u A_A-\epsilon (Y^C\nabla_CA_A+\frac{1}{2}\nabla_CY^C A_A+\frac{1}{2}(\nabla_AY_C-\nabla_CY_A)A^C).\nn\\ \label{ALC2}
\eea In the second line, we used the definition of CKV 
\be 
\nabla_AY_B+\nabla_BY_A-\frac{2}{d-2}\gamma_{AB}\nabla_CY^C =0.
\ee Comparing \eqref{CDA} and \eqref{ALC2}, we  obtain the same identification \eqref{CDLC}.

\paragraph{Definition of Carrollian amplitude.} In Lorenz gauge, the mode expansion of the vector field is \bea 
a_\mu(x)=\int\frac{d^{d-1}\bm p}{\sqrt{(2\pi)^{d-1}}}\frac{1}{\sqrt{2\omega_{\bm p}}}(\epsilon_{\mu}^a(\bm p) b_{a,\bm p}e^{-i\omega t+i\bm p\cdot\bm x}+\epsilon_\mu^a(\bm p) c_{a,\bm p}^\dagger e^{i\omega t-i\bm p\cdot\bm x})
\eea where the annihilation and creation operators satisfy the commutation relations 
\bea 
[b_{a,\bm p},b_{b,\bm p'}]=[c_{a,\bm p}^\dagger,c_{b,\bm p'}^\dagger]=0,\quad [b_{a,\bm p},c_{b,\bm p'}^\dagger]=\gamma_{ab}\delta(\bm p-\bm p').
\eea In this case, $\gamma_{ab}$ is the flat metric of $d-2$ dimensions and the polarization vector $\epsilon_\mu^a(\bm p)$ has $d-2$ transverse modes 
\be 
a=1,2,\cdots,d-2.
\ee 
The complex conjugate of $\epsilon_\mu^a$ is still a linear superposition of the same polarization vectors 
\be 
\left(\epsilon_\mu^a\right)^*=C^a_{\ b}\epsilon_\mu^b,
\ee where $C^a_{\ b}$ is a $(d-2)\times(d-2)$ constant matrix. From the reality condition of the vector field, we find the relation 
\bea 
c^\dagger_{a,\bm p}=C^b_{\ a}b^\dagger_{b,\bm p},\quad b_{a,\bm p}=C^b_{\ a}c_{b,\bm p}.
\eea 
The mode expansion of the boundary field $A_A(u,\Omega)$ is
\bea 
A_A(u,\Omega)&=&-Y^\mu_A(\Omega)\int_0^\infty \frac{d\omega}{\sqrt{4\pi\omega}}\omega^{\Delta}[e^{-i\pi \Delta/2}\epsilon_\mu^a(\bm p) b_{a,\bm p}e^{-i\omega u}+e^{i\pi \Delta/2}\epsilon_\mu^a(\bm p)c_{a,\bm p}^\dagger e^{i\omega u}]\nn\\&=&\int_0^\infty \frac{d\omega}{\sqrt{4\pi\omega}}\omega^{\Delta}[e^{-i\pi \Delta/2}e_A^a(\Omega) b_{a,\bm p}e^{-i\omega u}+e^{i\pi \Delta/2}e_A^a(\Omega)c_{a,\bm p}^\dagger e^{i\omega u}],
\eea  where we have defined the vielbein field 
\be 
e_A^a(\Omega)=-Y^\mu_A(\Omega)\epsilon_\mu^a(\bm p).
\ee 
The vector field in the local Cartesian frame is \be 
A_a(u,\Omega)=e_a^A(\Omega)A_A(u,\Omega)=\int_0^\infty \frac{d\omega}{\sqrt{4\pi\omega}}\omega^{\Delta}[e^{-i\pi \Delta/2} b_{a,\bm p}e^{-i\omega u}+e^{i\pi \Delta/2}c_{a,\bm p}^\dagger e^{i\omega u}].
\ee 
We can define the boundary state at $\mathcal{I}^+$
\bea 
|A_a(u,\Omega,+)\rangle=A_a(u,\Omega)|0\rangle=e^{i\pi \Delta/2}\int_0^\infty \frac{d\omega}{\sqrt{(2\pi)^{d-1}\times 8\pi}}\omega^{\Delta-1}e^{i\omega u}|a,\bm p\rangle,
\eea where the  state $|a,\bm p\rangle$ with definite momentum $\bm p$ and helicity $a$ is 
\be 
|a,\bm p\rangle=\sqrt{(2\pi)^{d-1}2\omega_{\bm p}}\ c_{a,\bm p}^\dagger |0\rangle.
\ee 
As  an outgoing state, it is better to define the following conjugate state
\bea 
\langle A_a(u,\Omega,+)|=\langle 0|A_a(u,\Omega)=e^{-i\pi \Delta/2}\int_0^\infty \frac{d\omega}{\sqrt{(2\pi)^{d-1}\times 8\pi}}\omega^{\Delta-1}e^{-i\omega u}\langle a,\bm p|
\eea where 
\be 
\langle a,\bm p|=\langle 0|b_{a,\bm p}\sqrt{(2\pi)^{d-1}2\omega_{\bm p}}.
\ee 
Now, we can also find the boundary field at $\mathcal{I}^-$
\bea 
A_A(v,\Omega)=-Y^\mu_A(\Omega)A_\mu(v,\Omega)
\eea with 
\be 
A_\mu(v,\Omega)=\int_0^\infty \frac{d\omega}{\sqrt{4\pi\omega}}\omega^{\Delta}[e^{i\pi \Delta/2}\epsilon_\mu^a(\bm p^{\text{P}})b_{{a,}\bm p^{\text{P}}}e^{-i\omega v}+e^{-i\pi \Delta/2}\epsilon_\mu^a(\bm p^{\text{P}})c^\dagger_{{ a,}\bm p^{\text{P}}}e^{i\omega v}].
\ee By imposing the antipodal map, 
\bea 
A_A^{\text{P}}(u,\Omega)&=&-\left(Y^{\text{P}}\right)^\mu_A(\Omega)A^{\text{P}}_\mu( u,\Omega)=Y^\mu_A(\Omega)A_\mu(v\to u,\Omega^{\text{P}})\nn\\&=&-\int_0^\infty \frac{d\omega}{\sqrt{4\pi\omega}}\omega^{\Delta}[e^{i\pi \Delta/2}e_A^a(\Omega)b_{a,\bm p}e^{-i\omega u}+e^{-i\pi \Delta/2}e_A^a(\Omega)c^\dagger_{a,\bm p}e^{i\omega u}].
\eea 
The vector field in the local Cartesian frame is \bea 
A^{\text{P}}_a(u,\Omega)=e_a^A(\Omega)A_A^{\text{P}}(u,\Omega)=-\int_0^\infty \frac{d\omega}{\sqrt{4\pi\omega}}\omega^{\Delta}[e^{i\pi \Delta/2}b_{a,\bm p}e^{-i\omega u}+e^{-i\pi \Delta/2}c^\dagger_{a,\bm p}e^{i\omega u}].
\eea 
Therefore, we can define the incoming state
\bea 
|A_a(u,\Omega,-)\rangle=A_a^{\text{P}}(u,\Omega)|0\rangle=-e^{-i\pi\Delta/2}\int_0^\infty \frac{d\omega}{\sqrt{(2\pi)^{d-1}\times 8\pi}}\omega^{\Delta-1}e^{i\omega u}|a,\bm p\rangle.
\eea 
Now we can define the vector Carrollian amplitude in higher dimensions
\bea 
&&i\, \mathcal{C}_{a_1a_2\cdots a_n}(u_1,\Omega_1,\sigma_1;\cdots;u_n,\Omega_n,\sigma_n)=\langle \prod_{j=1}^n A_{a_j}(u_j,\Omega_j,\sigma_j)\rangle\nn\\&=&\text{sign}_n\left(\frac{(-1)^{\Delta}}{(2\pi)^{d-1}\times 8\pi}\right)^{n/2}\prod_{j=1}^n \int_0^\infty d\omega_j \omega_j^{\Delta-1}e^{-i\sigma_j\omega_j u_j}(2\pi)^{d}\delta^{(d)}(\sum_j p_j)i\mathcal{M}_{a_1a_2\cdots a_n}(p_1,p_2,\cdots, p_n),\nn\\\label{CarrA}
\eea where 
\bea 
\mathcal{M}_{a_1a_2\cdots a_n}(p_1,p_2,\cdots,p_n)=\mathcal{M}_{\mu_1\mu_2\cdots\mu_n}(p_1,p_2,\cdots,p_n)\epsilon^{\mu_1}_{a_1}(\bm p_1)\cdots \epsilon^{\mu_n}_{a_n}(\bm p_n).
\eea 
The spacetime translation law of the Carrollian amplitude is 
\bea 
\mathcal{C}_{a_1a_2\cdots a_n}(u'_1,\Omega'_1,\sigma_1;\cdots;u'_n,\Omega'_n,\sigma_n)=\mathcal{C}_{a_1a_2\cdots a_n}(u_1,\Omega_1,\sigma_1;\cdots;u_n,\Omega_n,\sigma_n)
\eea and the Lorentz transformation law of the Carrollian amplitude is 
\bea 
\mathcal{C}_{a_1a_2\cdots a_n}(u'_1,\Omega'_1,\sigma_1;\cdots;u'_n,\Omega'_n,\sigma_n)=\left(\prod_{j=1}^n \Gamma_j^\Delta R_{a_j}^{\ b_j}\right)\mathcal{C}_{b_1b_2\cdots b_n}(u_1,\Omega_1,\sigma_1;\cdots;u_n,\Omega_n,\sigma_n).\label{Card}
\eea 
Now we will compute the boundary-to-boundary propagator 
\bea 
\langle A_a(u_1,\Omega_1,+)A_b(u_2,\Omega_2,-)\rangle=\frac{e^{-i\pi\Delta}}{4\pi}I_0(\omega_0(u_1-u_2-i\epsilon))\gamma_{ab}\delta(\Omega_1-\Omega_2).
\eea The bulk-to-boundary propagators are 
\bs\begin{align}
    \langle A_a(u,\Omega,+)a_\mu(x)\rangle&=\frac{e^{-i\pi\Delta} \Gamma(\Delta)}{4\pi\times (2\pi)^\Delta}\frac{\epsilon_{\mu a}(\Omega)}{(u+n\cdot x-i\epsilon)^\Delta},\\
    \langle a_\mu(x)A_a(u,\Omega,-)\rangle&=-\frac{ \Gamma(\Delta)}{4\pi\times (2\pi)^\Delta}\frac{\epsilon_{\mu a}(\Omega)}{(u+n\cdot x+i\epsilon)^\Delta}.\label{cav}
\end{align}\es 
 

\section{Superduality transformation}\label{sd}
In this section, we will discuss more on  the local rotation in the spinning Carrollian amplitude of general dimensions \eqref{Card} and elaborate on the emergence of superduality transformation of the spinning Carrollian field.   Assuming the Lorentz transformation is infinitesimal, the rotation matrix $R^{a}_{\ b}$ may be expanded as 
\bea 
R^a_{\ b}=\delta^a_{b}+\epsilon\Omega^a_{\ b},\quad a,b=1,2,\cdots,d-2,
\eea where $\Omega^a_{\ b}$ is the generator of the local rotation which is antisymmetric in the sense of
\be 
\Omega^{ab}=-\Omega^{ba}
\ee with 
\be 
\Omega^{ab}=\gamma^{cb}\Omega^a_{\ c}.
\ee To determine the form of the generator $\Omega^{a}_{\ b}$, we may compute $\delta A^a$ in two different ways. In the first approach, we use the finite transformation \eqref{transALC}
\bea 
\delta A^a(u,\Omega)&=&A'^a(u,\Omega)-A^a(u,\Omega)\nn\\&=&(1+\epsilon K)^\Delta (\delta^a_b+\epsilon\Omega^a_{\ b})(A^b-\epsilon Y^D\partial_D A^b+\epsilon {u}K \partial_u A^b)-A^a\nn\\&=&-\epsilon u\frac{\nabla_CY^C}{d-2}\partial_u A^a+\epsilon\Omega^a_{\ b}A^b-\epsilon Y^C\partial_C A^a-\epsilon\frac{1}{2}\nabla_CY^C A^a. \label{deltaAa}
\eea In the second approach, we use the variation of the field $A_A$ under Lorentz transformation 
\bea 
\delta A^a(u,\Omega)&=&\delta (e^a_A A^A)\nn\\&=&(\delta e^a_A)A^A+e^a_A \delta A^A\nn\\&=&(\delta e^a_A)A^A-\epsilon e^a_A(Y^C\nabla_C A^A+\frac{1}{2}\nabla_CY^C A^A+\frac{1}{2}(\nabla^AY^C-\nabla^CY^A)A_C+u\frac{\nabla_CY^C}{d-2}\partial_u A^A)\nn\\&=&(\delta e^a_A)A^A-\epsilon u\frac{\nabla_CY^C}{d-2}\partial_u A^a-\epsilon\frac{1}{2}\nabla_CY^C A^a-\epsilon e^a_A(Y^C\partial_CA^A+Y^C\Gamma^A_{CD}A^D\nn\\&&+\frac{1}{2}(\nabla^AY^C-\nabla^CY^A)A_C)\nn\\&=&(\delta e^a_A)A^A-\epsilon u\frac{\nabla_CY^C}{d-2}\partial_u A^a-\epsilon\frac{1}{2}\nabla_CY^C A^a-\epsilon Y^C\partial_CA^a\nn\\&&+\epsilon Y^C\partial_C e^a_A A^A-\epsilon e^a_A Y^CA^D\Gamma^A_{CD}-\epsilon e^a_A \frac{1}{2}(\nabla^AY^C-\nabla^CY^A)A_C\nn\\&=&(\delta e^a_A) A^A-\epsilon u\frac{\nabla_CY^C}{d-2}\partial_u A^a-\epsilon\frac{1}{2}\nabla_CY^C A^a-\epsilon Y^C\partial_CA^a-\epsilon Y^C  \omega_C^{\ ab}A_b\nn\\&&-\epsilon \frac{1}{2}e^a_A e_C^c (\nabla^AY^C-\nabla^CY^A)A_c.\label{deltaAa2}
\eea In the last step, we have introduced the spin connection $\omega_A^{\ ab}$ through the equation 
\bea 
\partial_Ce_A^{\ a}-\Gamma^D_{CA}e_D^a+\omega_{C}^{\ ab}e_{Ab}=0.\label{spinconnection}
\eea 
{In general, the variation of the vielbein field is a local rotation to preserve the invariance of the metric $\gamma_{AB}$.} Note that the metric $\gamma_{AB}$ is actually invariant under Lorentz transformations, the vielbein after the Lorentz transformation should be the same as {the} original one up to a local rotation 
\be 
\delta e^a_A=\epsilon\bar{\Omega}^a_{\ b}e^b_A,\label{vae}
\ee where $\bar{\Omega}^a_{\ b}$ is still antisymmetric 
\be 
\bar\Omega^{ab}=-\bar\Omega^{ba}.
\ee We have added a bar to distinguish it from the $\Omega^a_{\ b}$ associated with the matrix $R^a_{\ b}$. To match \eqref{deltaAa} with \eqref{deltaAa2}, we find 
\be 
\Omega^{ab}=\bar\Omega^{ab}-Y^C\omega_{C}^{\ ab}-\frac{1}{2}e_A^ae_B^b(\nabla^AY^B-\nabla^BY^A).\label{barOmega}
\ee 
In Appendix \ref{Local} where we also show that the last equation can be written as 
\bea 
\Omega^{ab}=\bar\Omega^{ab}+\Omega^{ab}_Y,
\eea where 
\bea 
\Omega_Y^{ab}&=&-\frac{1}{2}(\partial^aY^b-\partial^bY^a)-(\omega^{cab}+\frac{1}{2}(\omega^{abc}-\omega^{bac}))Y_c\nn\\&=&-\frac{1}{2}(D^aY^b-D^bY^a)-Y_c\omega^{cab}  \label{Omega}
\eea with $\partial^a$ 
\be 
\partial^a=e^{Aa}\partial_A
\ee and $Y^a$ the CKV in the local frame
\be 
Y^a=e_A^a Y^A.
\ee The covariant derivative of $Y^a$ is 
\be 
D_aY^b=\partial_a Y^b+\omega_a^{\ bc}Y_c.
\ee 

After fixing the vielbein field, the spin connection is determined unambiguously  and the last three terms of \eqref{barOmega}  are completely fixed consequently. However, the matrix $\bar{\Omega}^{ab}$ is freely chosen because it is related to the local $SO(d-2)$ rotation. Usually, in the computation of the scattering amplitude, the polarization vectors are chosen to take the same form after Lorentz transformations, as has been presented in \eqref{inve}
\be 
e'^a_A(\Omega')=e^a_A(\Omega').\label{vielbeininv}
\ee For instance, in the previous examples of Carrollian amplitude, the form of the vielbein field is always  chosen as \eqref{fixvielbein}. It follows from \eqref{vielbeininv} that the $\bar{\Omega}^{ab}$ vanishes 
\be 
\bar\Omega^{ab}=0
\ee and $\Omega^{ab}$ is completely fixed by the vector $Y^a$ of the Lorentz transformations  in this convention.

However, there is an ambiguity  in the choice of the  local frame for the vielbein field which is described by the matrix $\bar\Omega^{ab}$.  To discuss its effect, we set $Y^a=0$ and then the variation of the field $A^a$ is totally determined by $\bar\Omega^{ab}$
\be 
\delta A^a=\epsilon \bar\Omega^{ab}A_b.\label{deltaAabarOmega}
\ee 

We notice that there is a similar local rotation in \cite{Liu:2023qtr} which is to rotate the boundary field $A_A$ and its Hodge dual $\epsilon_{AB}A^B$ in four dimensions. More explicitly, this is realized by the so-called helicity flux operator 
\bea 
\mathcal{O}_g=\int du d\Omega g(\Omega) \epsilon_{AB}\dot{A}^B A^A
\eea whose commutator with the vector field $A_A$ is 
\bea 
[\mathcal{O}_g, A_A(u,\Omega)]=i g(\Omega)\epsilon_{AB}A^B(u,\Omega).
\eea In terms of the vector field in the local frame, this is equivalent to
\be 
[\mathcal{O}_g, A_a(u,\Omega)]=i g(\Omega)\epsilon_{ab}A^b(u,\Omega)\label{deltaAabarOmega2}
\ee where we have used the identity 
\be 
\epsilon_{ab}=e_a^A\epsilon_{AB}\epsilon_b^B.
\ee The variation \eqref{deltaAabarOmega} matches with \eqref{deltaAabarOmega2} provided 
\be 
\bar\Omega^{ab}=g(\Omega) \epsilon^{ab}.
\ee 
In four dimensions, the celestial sphere is two-dimensional and then  $\bar\Omega^{ab}$ should be proportional to $\epsilon^{ab}$. So the antisymmetric matrix $\bar\Omega^{ab}$ is in one-to-one correspondence with the function $g(\Omega)$. In higher dimensions, there is a similar helicity flux operator \cite{Liu:2024rvz}
\be 
\mathcal{O}_{\bm h}=\int du d\Omega h_{AB}(\Omega)\dot{A}^BA^A,\quad h_{AB}=-h_{BA}
\ee whose commutator with $A_a$ is 
\bea 
[\mathcal{O}_{\bm h},A_a]=ih_{ab}A^b,\quad h_{ab}=e^A_a e^B_b h_{AB}.\label{OAa}
\eea Interestingly, since both $h_{ab}$ and $\bar\Omega_{ab}$ are antisymmetric, there is still a one-to-one correspondence between them 
\be 
\bar\Omega_{ab}=h_{ab}.
\ee 

The superduality transformation $S$ is a local $SO(2)$ rotation of the vector field which is generated by the helicity flux operator \cite{Liu:2023qtr,Liu:2023gwa,Liu:2023jnc,Liu:2024nkc}
\be 
S: A_a\to S_a^{\ b}A_b.
\ee  In higher dimensions, it is extended to a local $SO(d-2)$ rotation of the vector field, even though there is no similar electromagnetic  duality invariance for higher dimensional gauge theories.
Obviously, the correspondence between $\bar\Omega_{ab}$ and $h_{ab}$ build an isomorphism between the local rotation of the vielbein field and the superduality transformation. 

Before we close this section, we should comment on this intriguing isomorphism. Indeed, the variation \eqref{deltaAabarOmega} has the same form as the variation \eqref{OAa}. However, there is still a distinction. On the one hand, the variation  \eqref{deltaAabarOmega} is obtained by fixing $A_A$ and rotating the local frame $e_a^A$.  On the other hand, the variation \eqref{OAa} is found by fixing the frame $e_a^A$ while rotating the field $A_A$ at the same time.  The former one is a passive view that changes the description without altering the physical system. The latter one is a positive view that rotates the state $|A_a(u,\Omega)\rangle$ to another state, which indeed changes the physical system.

\section{Discussion and conclusion}\label{conc}
By definition, Carrollian amplitude is a geometric object which lives on the Carrollian manifold. Motivated by this idea,  we work out the proper definition of the  Carrollian amplitude in the framework of bulk reduction.  For any radiative field $F_{A(s)}$ at the null infinity, we may use the vielbein field $e^{A}_{a}$ on the unit sphere to obtain the fundamental field 
\be 
F_{a(s)}=F_{A(s)}e^{A(s)}_{a(s)},\quad e^{A(s)}_{a(s)}=e_{a_1}^{A_1}\cdots e_{a_s}^{A_s}
\ee in the local Cartesian frame. A general Carrollian amplitude is the correlation function of the fundamental field 
\bea 
i\, \mathcal{C}_{a_1(s_1)\cdots a_n(s_n)}=\langle \prod_{j=1}^n F_{a_j(s_j)}(u_j,\Omega_j,\sigma_j)\rangle
\eea which is related to the standard scattering amplitude by modified Fourier transform 
\bea 
i\, \mathcal{C}_{a_1(s_1)\cdots a_n(s_n)}=\prod_{j=1}^n \int_0^\infty d\omega_j \omega_j^{\Delta-1}e^{-i\sigma_j \omega_j u_j}i \mathcal{M}_{a_1(s_1)\cdots a_n(s_n)}(p_1,p_2,
\cdots,p_n)\label{relation}
\eea where $\Delta=\frac{d-2}{2}$. This definition is reduced to the Fourier transform in four dimensions and independent  of the coordinates used on the unit sphere $S^{d-2}$.
It is also consistent with the previous work on the Carrollian amplitude of gluons or gravitons in spinor helicity formalism. The relation \eqref{relation} between the Carrollian amplitude and the scattering amplitude still holds for fermionic theories. To make this statement robust, one should extend the discussion to fermionic theory and discuss the asymptotic expansion of the Dirac field. We expect to present the relevant work in the near future.

We have derived the Poincar\'e transformation of the Carrollian amplitude for  spinning theory in general dimensions. There are two important distinctions compared to the scalar Carrollian amplitude of four dimensions. At first, there could be a rotation matrix $R_{a(s)}^{\ b(s)}$ related to the Lorentz transformation for general spin theory. Second, the redshift factor $\Gamma$ is dressed with a weight $\Delta$ in general dimensions. Moreover, it becomes clear that the superduality transformation which rotates the fundamental field is isomorphic to the frame choice of the vielbein field. Note that superduality transformation always appears even though there is no similar  electromagnetic duality transformation in general dimensions and there is always a helicity flux operator to generate superduality transformation. The helicity flux operator may cause the spin precession of freely falling gyroscopes \cite{Seraj:2022qyt} and the angular distribution of  helicity flux density from binary systems has been explored in \cite{Dong:2024ily}. It is interesting to understand the physical consequences of the helicity flux operator in the future.

 Our work shows clearly that Carrollian amplitude is an object that integrates quantum effects, geometry and holographic principle nicely. In Table \ref{table1}, we compare the AdS/CFT and Carrollian holography in parallel.
      \begin{table}[h]
    \centering
    \begin{tabular}{|c|c|}  \hline\text{AdS/CFT}&\text{Carrollian holography}\\\hline\hline
   $d\to d-1$&$d\to d-1$\\\hline
   Fefferman Graham expansion&Asymptotic expansion\\\hline
   \text{AdS scattering}&\text{Flat space scattering}\ \\\hline
   \text{Witten diagram}&\text{Feynman diagram}\\\hline
    \end{tabular}
   \caption{\centering{Analogy between AdS/CFT and Carrollian holography.}}
    \label{table1}
\end{table}
Let us explain the similarity in more details. At first, both AdS/CFT and Carrollian holography state that a $d$-dimensional gravitational theory is equivalent to a $(d-1)$-dimensional quantum field theory which lives on its boundary. To relate the bulk field and the boundary field, one can use Fefferman Graham expansion \cite{AST_1985__S131__95_0} in AdS/CFT and asymptotic expansion in Carrollian holography. The asymptotic expansion is one of the ingredients of bulk reduction which is used in this work. In AdS/CFT, one uses AdS scattering to determine the CFT correlators holographically. This is realized by Witten diagrams. In Carrollian holography, one should use flat space scattering to calculate the Carrollian amplitude which is the holographic dual of boundary correlators. The diagrammatic way to calculate the Carrollian amplitude is the usual Feynman rules in the bulk and the bulk-to-boundary propagator which is to define the external lines \cite{Liu:2024nfc}. Moreover, there is a smooth limit from AdS holography to Carrollian holography in the AdS Bondi-coordinates\cite{Barnich:2012aw,Poole:2018koa,Compere:2019bua,Compere:2020lrt}, as has been shown recently \cite{Alday:2024yyj}. We hope this quantity may shed new light on the longstanding problem of quantum gravity.

\vspace{10pt}
{\noindent \bf Acknowledgments.}
The work of J.L. was supported by NSFC Grant No. 12005069. The work of W.-B. Liu, H.-Y. Xiao and J.-L. Yang is supported by ``the Fundamental Research Funds for the Central Universities'' with
No. YCJJ20242112.

\appendix
\section{Conformal Killing vector, polarization vector and vielbein field}\label{vielbeinapp}
In this Appendix, we will discuss the relation between the polarization vector and the vielbein field on the celestial sphere. We will work on general $d$ dimensions and the sphere is $d-2$
dimensional. The unit normal vector of the $S^{d-2}$ is $n^i$ and we may construct the following conformal Killing vectors
\be 
Y^i_A=-\nabla_A n^i.
\ee 
The vectors $Y^i_A$ are lifted to $d$ dimensions 
\be 
Y^\mu_A=-\nabla_A n^\mu
\ee which are orthogonal to $n^\mu$ and $\bar n^\mu$
\be 
Y^\mu_A n_\mu=Y^\mu_A \bar n_\mu=0
\ee and obey the identities 
\be 
Y^\mu_A Y^\nu_B \gamma_{\mu\nu}=\gamma_{AB},\quad Y^\mu_A Y^\nu_B\gamma^{AB}=\gamma^{\mu\nu}.
\ee These identities are similar to those for polarization vectors \eqref{orthocomp}. By contracting the Lorentz indices, we can define the vielbein field for the vector field 
\be 
e_A^a=-Y_A^\mu\epsilon_\mu^a.
\ee At first, this is an exactly $(d-2)\times (d-2)$ matrix. Secondly, we can check the orthogonal and completeness relations 
\bs\begin{align}
    e_A^a e_B^b \gamma^{AB}&=Y^\mu_A \epsilon_\mu^a Y^\nu_B\epsilon_\nu^b\gamma^{AB}=\epsilon_\mu^a\epsilon_\nu^b \gamma^{\mu\nu}=\gamma^{ab},\\
    e_A^a e_B^b \gamma_{ab}&=Y^\mu_A \epsilon_\mu^a Y^\nu_B\epsilon_\nu^b\gamma_{ab}=Y^\mu_A Y^\nu_B \gamma_{\mu\nu}=\gamma_{AB}.
\end{align}\es 

Now we will construct the polarization tensor for gravitons 
\be 
\epsilon_{\mu\nu}^{ab}=\frac{1}{2}(\epsilon_\mu^a\epsilon_\nu^b+\epsilon_\nu^a\epsilon_\mu^b-\frac{2}{d-2}\gamma^{ab}\gamma_{\mu\nu}).
\ee We can check that this symmetric polarization tensor is transverse and traceless 
\bs\begin{align}
\epsilon_{\mu\nu}^{ab}n^\mu&=0,\\
\epsilon_{\mu\nu}^{ab}\gamma_{ab}&=\epsilon_{\mu\nu}^{ab}\gamma^{\mu\nu}=0.
\end{align}\es To check the transverse condition, we have used the identity 
\be 
\epsilon_\mu^a n^\mu=0,\quad \gamma_{\mu\nu}n^\mu=0.
\ee The traceless condition follows from \eqref{orthocomp} and the identity 
\be 
\gamma_{ab}\gamma^{ab}=\gamma_{\mu\nu}\gamma^{\mu\nu}=d-2
\ee The number of independent combinations  of symmetric and traceless indices $(ab)$ is
\be 
\frac{(d-2)(d-1)}{2}-1=\frac{d(d-3)}{2}
\ee which is exactly the number of propagating degrees of freedom of gravitons.
\section{Ingredients on spinor helicity formalism}\label{sh}
In this appendix, we will review the basic ingredients to connect the definition to the usual scattering amplitude. We will work in the spinor helicity formalism which is powerful for massless particles. 

\paragraph{Spinor helicity formalism.} In this formalism, we will write $u_s(\bm p),\ v_s(\bm p)$ of massless electrons and positrons with definite helicity $s$ and momentum $\bm p$ as follows \cite{2007qft..book.....S,2015sagt.book.....E}
\bs\begin{align}
    &|p]=u_-(\bm p)=v_+(\bm p)=\left(\begin{array}{c}\lambda_a\\ 0\end{array}\right),\\
    &|p\rangle=u_+(\bm p)=v_-(\bm p)=\left(\begin{array}{c}0\\\lambda^{*\dot{a}}\end{array}\right),\\
    &[p|=\bar{u}_+(\bm p)=\bar{v}_-(\bm p)=(\lambda^a,0),\\
    &\langle p|=\bar{u}_-(\bm p)=\bar{v}_+(\bm p)=(0,\lambda^{*}_{\dot a}).
\end{align}\es For real momentum, we have 
\be 
\lambda^*_{\dot a}=(\lambda_a)^*.
\ee 
The lower index of the commuting spinors $\lambda_a,\ \lambda_{\dot a}^*$ are raised by the invariants of the Lorentz group 
\bea 
\lambda^a=\epsilon^{ab}\lambda_b,\quad \lambda^{*\dot a}=\epsilon^{\dot a\dot b}\lambda_{\dot b}^*
\eea with 
\bea 
\epsilon_{ab}=\epsilon_{\dot a\dot b}=\left(\begin{array}{cc}0&-1\\1&0\end{array}\right),\quad \epsilon^{ab}=\epsilon^{\dot a\dot b}=\left(\begin{array}{cc}0&1\\-1&0\end{array}\right).
\eea The form of $\lambda_a$ is only determined up to a phase, 
an explicit representation of $\lambda_a$ is \cite{2007qft..book.....S}
\be 
\lambda_a=\sqrt{2\omega}\left(\begin{array}{c}-\sin\frac{\theta}{2}e^{-i\phi}\\\cos\frac{\theta}{2}\end{array}\right)
\ee where $\theta,\phi$ are the polar and azimuthal angles which represent the direction of the momentum $p$. In terms of stereographic coordinates
\be 
z=\cot\frac{\theta}{2}e^{i\phi},\quad \bar z=\cot\frac{\theta}{2}e^{-i\phi},
\ee this is equivalent to 
\be 
\lambda_a=\sqrt{2\omega}\left(\begin{array}{c}-\sqrt{\frac{\bar z}{z(1+z\bar z)}}\\ \sqrt{\frac{z\bar z}{1+z\bar z}}\end{array}\right).
\ee Note that ${\bar z}/{z}$
is just a phase, we will choose the following convention 
\bea 
\lambda_a=\sqrt{\frac{2\omega}{1+z\bar z}}\left(\begin{array}{c}-1\\ z\end{array}\right).
\eea As a consequence, 
\bs\begin{align}
     \lambda_{\dot a}^*&=\sqrt{\frac{2\omega}{1+|z|^2}}(-1,\bar z),\\
    \lambda^a&=\sqrt{\frac{2\omega}{1+|z|^2}}(z,1),\\
    \lambda^{*\dot a}&=\sqrt{\frac{2\omega}{1+|z|^2}}{ \left(\begin{array}{c}\bar z\\  1\end{array}\right)}.
\end{align}\es The null momentum $p^\mu$ may be transformed to a $2\times 2$ matrix
\be
    p_{a\dot a}=p^\mu\sigma_{\mu a\dot a}=\frac{2\sigma\omega}{1+z\bar z}\left(\begin{array}{cc}-1&\bar z\\ z&-z\bar z\end{array}\right)=-\sigma\lambda_a\lambda_{\dot a}^*,
\ee where we have used the four-momentum 
\be 
p^\mu=\sigma \omega n^\mu\label{4momentum}
\ee and the matrix 
\bea 
\sigma^\mu=(1,\sigma^i).
\eea Note that $\sigma^i$ are Pauli matrices 
\be 
\sigma^1=\left(\begin{array}{cc}0&1\\ 1&0\end{array}\right),\quad \sigma^2=\left(\begin{array}{cc}0&-i\\ i&0\end{array}\right),\quad \sigma^3=\left(\begin{array}{cc}1&0\\ 0&-1\end{array}\right).
\ee One should distinguish them with the symbol $\sigma$ which is used to represent incoming or outgoing states. We can also define 
\be 
\bar\sigma^{\mu\dot a a}=\epsilon^{ab}\epsilon^{\dot a\dot b}\sigma^\mu_{b\dot b}=(1,-\sigma^i)
\ee and obtain 
\be 
p^{\dot a a }=p^\mu\bar{\sigma}^{\dot a a}_\mu=-\sigma\lambda^{*\dot a}\lambda^a.
\ee It is straightforward to compute the following products 
\be 
\langle p_1p_2\rangle=\lambda_{(1)\dot a}^* \lambda_{(2)}^{*\dot a},\quad [p_1p_2]=\lambda_{(1)}^{a}\lambda_{(2)a}
\ee where $\lambda_{(1)a}$ and $\lambda_{(2)a}$ are the spinors associated with momentum $p_1$ and $p_2$, respectively. Sometimes we will omit the symbol of momenta 
\be 
\langle 12\rangle=\langle p_1p_2\rangle,\quad [12]=[p_1p_2].
\ee With the convention \eqref{4momentum}, we have 
\bs\begin{align}
    \langle12\rangle&=\sqrt{\frac{4\omega_1\omega_2}{(1+|z_1|^2)(1+|z_2|^2)}}\bar{z}_{12}=\sqrt{4w_1w_2}\bar{z}_{12},\\
    [12]&=-\sqrt{\frac{4\omega_1\omega_2}{(1+|z_1|^2)(1+|z_2|^2)}}z_{12}=-\sqrt{4w_1w_2}z_{12},\\
    \langle 12\rangle [12]&=-\frac{4\omega_1\omega_2}{(1+|z_1|^2)(1+|z_2|^2)}z_{12}\bar{z}_{12}=-4w_1w_2z_{12}\bar{z}_{12}=2\sigma_1\sigma_2 p_1\cdot p_2,
\end{align}\es where 
\be 
z_{12}=z_1-z_2,\quad \bar{z}_{12}=\bar{z}_1-\bar{z}_2
\ee and we have defined 
\be 
w_i=\frac{\omega_i}{1+|z_i|^2},\quad i=1,2.
\ee 

\paragraph{Lorentz transformation.}
The Lorentz transformation induces a conformal transformation of the unit sphere and it can be represented by a  M\"obius transformation in stereographic coordinates
\be 
z'=\frac{az+b}{cz+d},\quad ad-bc=1,\quad a,b,c,d\in\mathbb{C}.
\ee 
It follows that $z_{ij}=z_i-z_j$ is transformed to 
\be 
z_{ij}'=\frac{z_{ij}}{(cz_i+d)(cz_j+d)}.\label{transz}
\ee Moreover, 
as has been shown in \cite{Liu:2024nfc}, the frequency $\omega$ is redshifted under Lorentz transformation with the redshift factor 
\bea 
\Gamma=\frac{(1+|z'|^2)|cz+d|^2}{1+|z|^2}.
\eea Therefore, the Lorentz transformation  of $w=\frac{\omega}{1+|z|^2}$ is rather simple
\be 
w'=|cz+d|^2 w.
\ee Combining with \eqref{transz}, we find the following transformation law of the angle and square brackets 
\bea 
\langle 1'2'\rangle=(t_1^*t_2^*)^{1/2}\langle 12\rangle,\quad [1'2']=(t_1t_2)^{1/2}[12],\label{Lorentzbracket}
\eea where 
\be 
t_i={\frac{\bar c \bar z_{i}+\bar d}{cz_{i}+d}}=\left(\frac{\partial z_i'}{\partial z_i}\right)^{1/2}\left(\frac{\partial\bar z_i'}{\partial z_i}\right)^{-1/2},\quad t_i^*=t_i^{-1}.\label{littlet}
\ee We  have used the following identity when calculating the factor coming from Lorentz transformation in \eqref{gammat} 
\be 
\Gamma\ t=\frac{1+|z'|^2}{1+|z|^2}(\bar c\bar z+\bar d)^2,\quad \Gamma\ t^*=\frac{1+|z'|^2}{1+|z|^2}(c z+d)^2.\label{B27}
\ee 

\paragraph{Polarization vector.} 
In spinor helicity formalism, the polarization vectors may be chosen as 
\bea 
\epsilon_\mu^{-}(\bm p)=e^{2i\varphi}\frac{\langle q|\gamma_\mu|p]}{\sqrt{2}\langle qp\rangle},\quad 
\epsilon_\mu^{\text{+}}(\bm p)=-e^{-2i\varphi}\frac{[q|\gamma_\mu|p\rangle}{\sqrt{2}[qp]}
\eea where $\varphi$ is an arbitrary angle and $q$ is a massless reference momentum
\be 
q_{a\dot a}=-\sigma \kappa_a\kappa_{\dot a}^*.
\ee The Dirac matrix $\gamma^\mu$ is 
\be 
\gamma^\mu=\left(\begin{array}{cc}0&\sigma^\mu\\ \bar\sigma^\mu&0\end{array}\right).
\ee We may choose the reference momentum as 
\be 
q_{a\dot a}=-\frac{2\sigma \omega_q}{1+z\bar z}\left(\begin{array}{cc}z\bar z&\bar z\\ z&1\end{array}\right)
\ee and the phase 
\be 
\varphi=\phi.
\ee Then the polarization vectors are
 \bs\begin{align}
\epsilon_\mu^{-}&=\frac{\sqrt{2}}{1+z\bar z}\left(0,\frac{1-z^2}{2},-\frac{1+z^2}{2i},z\right),\\
\epsilon_\mu^{+}&=\frac{\sqrt{2}}{1+z\bar z}\left(0,\frac{1-\bar z^2}{2},\frac{1+\bar z^2}{2i},\bar z\right).
\end{align}\es  
The two polarization vectors are related to each other by complex conjugate 
\bea 
\left(\epsilon_\mu^+\right)^*=\epsilon_\mu^-.
\eea Recalling the polarization vectors in canonical quantization, we have the following correspondence 
\be 
1\leftrightarrow -,\quad 2\leftrightarrow +.
\ee To match with  \eqref{complexC}, we find the complex conjugate matrix 
\be 
C^a_{\ b}=\left(\begin{array}{cc}0&1\\ 1&0\end{array}\right).
\ee Considering the orthogonality and completeness relation \eqref{orthocomp}, the flat metric $\gamma^{ab}$ is 
\be 
\gamma^{ab}=\left(\begin{array}{cc}0&1\\ 1&0\end{array}\right),\quad \gamma_{ab}=\left(\begin{array}{cc}0&1\\ 1&0\end{array}\right).
\ee Indeed, as suggested in \eqref{Cab}, we find 
\be 
C_{ab}=\gamma_{ac}C^c_{\ b}=\delta_{ab}=\left(\begin{array}{cc}1&0\\ 0&1\end{array}\right).
\ee 
The vielbein field is 
\bea 
e_A^a=-Y_A^\mu\epsilon_\mu^a=\frac{\sqrt{2}}{1+z\bar z}\left(\begin{array}{cc}1&0\\ 0&1\end{array}\right),\label{fixvielbein}
\eea which leads to the Lorentz transformation law  \eqref{LorentzR} with
\be 
R^a_{\ b}=\left(\begin{array}{cc}t&0\\ 0&t^*\end{array}\right),\quad R_a^{\ b}=\left(\begin{array}{cc}t^*&0\\ 0&t\end{array}\right)\label{Rab}
\ee after imposing the condition \eqref{inve}.
\section{Local rotation in the Cartesian frame}\label{Local}
In this appendix, we will prove several statements in the context. At first, we will check the infinitesimal variation \eqref{vae}
\bea 
\delta e^a_A&=&e'^{a}_A-e^a_A\nn\\&=&(1-\epsilon K)(\delta^C_A-\epsilon\partial_AY^C)(\delta^a_{b}+\epsilon\Omega^a_{\ b})(e^b_C-\epsilon Y^D\partial_D e_C^b)-e^a_A\nn\\&=&\epsilon\frac{\nabla_CY^C}{d-2}e_A^a-\epsilon\partial_AY^C e_C^b+\epsilon\Omega^a_{\ b}e_A^b-\epsilon Y^D\partial_D e_A^a\nn\\&=&\epsilon \frac{\nabla_CY^C}{d-2}e_A^a-\epsilon\nabla_AY^C e_C^a+\epsilon\Omega^a_{\ b}e_A^b+\epsilon Y^D\omega_D^{\ ab}e_{Ab}\nn\\&=&\epsilon\Omega^a_{\ b}e_A^b+\epsilon Y^C\omega_{C}^{\ ab}e_{Ab}-\epsilon\frac{1}{2}(\nabla_AY_C-\nabla_CY_A)e^{Ca}\nn\\&=&\epsilon\bar{\Omega}^a_{\ b}e^b_A.
\eea In the second line, we have used the finite variation \eqref{LorentzR}. In the fourth line, we have used the definition spin connection \eqref{spinconnection}. In the fifth line, we have used the fact that $Y^A$ is a CKV. In the last line, we substitute the relation between $\Omega^{ab}$ and $\bar{\Omega}^{ab}$ \eqref{barOmega}.

Secondly, we will prove the formula \eqref{Omega}. It is sufficient to compute 
\bea 
&&Y^C\omega_{C}^{\ ab}+\frac{1}{2}e_A^ae_B^b(\nabla^AY^B-\nabla^BY^A)\nn\\&=&Y^c\omega_{c}^{\ ab}+\frac{1}{2}e^{Aa}e^{Bb}(\partial_AY_B-\partial_BY_A)\nn\\&=&Y^c\omega_{c}^{\ ab}+\frac{1}{2}e^{Aa}(\partial_A Y^b-Y_B\partial_A e^{Bb})-\frac{1}{2}e^{Bb}(\partial_B Y^a-Y_A\partial_B e^{Aa})\nn\\&=&\frac{1}{2}(\partial^aY^b-\partial^bY^a)+Y^c\omega_c^{\ ab}-\frac{1}{2}e^{Aa}Y_B(-\Gamma^{B}_{AD}e^{Db}-\omega_A^{\ bc}e^B_c)+\frac{1}{2}e^{Bb}Y_A(-\Gamma^A_{BD}e^{Da}-\omega_B^{\ ac}e^{A}_{c})\nn\\&=&\frac{1}{2}(\partial^aY^b-\partial^bY^a)+Y_c(\omega^{cab}+\frac{1}{2}\omega^{abc}-\frac{1}{2}\omega^{bac})\nn\\&=&\frac{1}{2}(D^aY^b-D^bY^a)+Y_c\omega^{cab}.
\eea Note that the spin connection is related to the structure constant by 
\be 
\omega^{cab}-\omega^{bac}=f^{cba}\label{omegaf}
\ee where structure constant is defined by the commutators of the vielbein field
\be 
[e^a,e^b]=f^{ab}_{\ \ c}e^c=f^{abc}e_c.
\ee More explicitly, 
\bea 
f^{abc}=e_A^c(e^{Ca}\partial_Ce^{Ab}-e^{Cb}\partial_C e^{Aa})=e_A^c[e^a,e^b]^A.
\eea 
Therefore, we find another form for $\Omega_Y^{ab}$
\bea 
-\Omega_Y^{ab}&=&Y^C\omega_{C}^{\ ab}+\frac{1}{2}e_A^ae_B^b(\nabla^AY^B-\nabla^BY^A)\nn\\&=&
\frac{1}{2}(\partial^aY^b-\partial^bY^a)+\frac{1}{2}(f^{cab}-f^{cba})Y_c.\label{omegaYformula1}
\eea Since the spin connection is antisymmetric for the last two indices, we can solve \eqref{omegaf} 
\be 
\omega^{abc}=-\frac{1}{2}(f^{abc}-f^{bca}+f^{cab}).
\ee 

As an example, we compute the structure constant associated with the vielbein field \eqref{fixvielbein}
\be 
e^{Aa}=\frac{1+z\bar z}{\sqrt{2}}\left(\begin{array}{cc}0&1\\1&0\end{array}\right),\quad e^A_a=\frac{1+z\bar z}{\sqrt{2}}\left(\begin{array}{cc}1&0\\0&1\end{array}\right).
\ee We can construct the following field 
\bea 
e^1=e_2=\frac{1+z\bar z}{\sqrt{2}}\partial_{\bar z},\quad e^2=e_1=\frac{1+z\bar z}{\sqrt{2}}\partial_{z}.
\eea Therefore, 
\bea 
[e^1,e^2]=\frac{z}{\sqrt{2}}e_1-\frac{\bar z}{\sqrt{2}}e_2\quad\Rightarrow\quad f^{121}=\frac{z}{\sqrt{2}},\quad f^{122}=-\frac{\bar z}{\sqrt{2}}.
\eea The spin connections are 
\bea 
\omega^{112}=-\omega^{121}= \frac{z}{\sqrt{2}},\quad \omega^{212}=-\omega^{221}=-\frac{\bar z}{\sqrt{2}}.
\eea 

In the local Cartesian frame, the CKVs are 
\be 
Y_{ia}=-e^A_a\nabla_A n_i,\quad Y_{ija}=Y_{ia}n_j-Y_{ja}n_i
\ee whose components are 
\bs\begin{align}
    Y_{1a}&=-\frac{1}{\sqrt{2}(1+z\bar z)}(1-\bar z^2,1-z^2),\quad Y_1^a=-\frac{1}{\sqrt{2}(1+z\bar z)}(1-z^2,1-\bar z^2),\\
    Y_{2a}&=\frac{i}{\sqrt{2}(1+z\bar z)}(1+\bar z^2,-1-z^2),\quad Y_2^a=\frac{i}{\sqrt{2}(1+z\bar z)}(-1-z^2,1+\bar z^2),\\
    Y_{3a}&=-\frac{\sqrt{2}}{1+z\bar z}(\bar z,z),\quad Y_3^a=-\frac{\sqrt{2}}{1+z\bar z}(z,\bar z),\\
    Y_{12a}&=\frac{i\sqrt{2}}{1+z\bar z}(-\bar z,z),\quad Y_{12}^a=\frac{i\sqrt{2}}{1+z\bar z}(z,-\bar z),\\
    Y_{23a}&=\frac{i}{\sqrt{2}(1+z\bar z)}(-1+\bar z^2,1-z^2),\quad Y_{23}^a=\frac{i}{\sqrt{2}(1+z\bar z)}(1-z^2,-1+\bar z^2),\\
    Y_{13a}&=\frac{1}{\sqrt{2}(1+z\bar z)}(1+\bar z^2,1+z^2),\quad Y_{13}^a=\frac{1}{\sqrt{2}(1+z\bar z)}(1+ z^2,1+\bar z^2).
\end{align}\es

In  Table \ref{omegaY1}, we compute the $\Omega_Y=-\frac{1}{2}\epsilon_{ab}\Omega_Y^{ab}$ using formula \eqref{omegaYformula1}.
 \begin{table}[h]
    \centering
    \begin{tabular}{|c|c|c|}  \hline\text{CKV}&\text{$\Omega_Y$}&\text{Infinitesimal M\"obius transformations}\\\hline\hline
   $Y_1$&$\frac{z-\bar z}{2}$&$z'=z+\epsilon \frac{z^2-1}{2}$\\\hline
   $Y_2$&$-\frac{i(z+\bar z)}{2}$&$z'=z-\epsilon \frac{i(1+z^2)}{2}$\\\hline
   $Y_3$&$0$ &$z'=z-\epsilon z$\\\hline
   $Y_{12}$&$i$&$z'=z+i\epsilon z$\\\hline
   $Y_{23}$&$-\frac{i(z+\bar z)}{2}$&$z'=z+\epsilon\frac{i(1-z^2)}{2}$\\\hline
   $Y_{13}$&$\frac{z-\bar z}{2}$&$z'=z+\epsilon\frac{1+z^2}{2}$\\\hline
    \end{tabular}
  \caption{\centering{$\Omega_Y$ {and the corresponding infinite M\"obius transformations for CKVs\\ in 4 dimensions.}}}
    \label{omegaY1}
\end{table}
Note that with the convention in the context, the matrix $R^a_{\ b}$ is also determined by the phase
\be 
t=\left(\frac{\partial z'}{\partial z}\right)^{1/2}\left(\frac{\partial \bar z'}{\partial \bar z}\right)^{-1/2}.
\ee For a general conformal transformation 
\be z'=z+\epsilon f(z),
\ee we can find 
\bea 
t=1+\epsilon \frac{1}{2}(f'(z)-\bar{f}'(\bar z))\quad\Rightarrow\quad \Omega_Y=\frac{1}{2}(f'(z)-\bar{f}'(\bar z)).
\eea One can check that the $\Omega_Y$ in the second column is consistent with the M\"obius transformation in the third column. Therefore, we have checked that the generator  \eqref{omegaYformula1} indeed generates the rotation matrix $R^a_{\ b}$ in four-dimensional theories.

\section{Comparison}\label{comparison}
To compare with the result in \cite{Mason:2023mti}, one should notice the following three differences  in the conventions.
\begin{enumerate}
    \item The role of angle and square brackets are exactly opposite. In other words, the right-hand side of \eqref{pmmp} in \cite{Mason:2023mti} is actually the amplitude \bea 
    \mathcal{M}^{+,-,-,+}[1,2,3,4]=\frac{\omega_2\omega_3}{\omega_1\omega_4}\frac{1+|z|^2}{2\bar z}\label{4gluon}
    \eea in our convention.
    \item The cross ratio in the paper \cite{Mason:2023mti} is $Z=\frac{z_{12}z_{34}}{z_{13}z_{24}}$ while the cross ratio in this paper is $z=-\frac{z_{14}z_{23}}{z_{12}z_{34}}$.
    \item The convention of the four-momentum $p^\mu$ is also different.
\end{enumerate}

\paragraph{MHV amplitude for four gluons.}
In \cite{Mason:2023mti}, the Carrollian amplitude for four gluons is proportional to \begin{align}
    \frac{(1-Z)\bar z_{14}^{2}}{Z z_{14}^{2}\bar z_{13}^{2} \bar z_{24}^{2}}\delta(Z-\bar Z)&=- \frac{z}{\bar z_{13}^{2} \bar z_{24}^{2}} \frac{\bar z_{14}^{2}}{z_{14}^{2}}(1-z)^{2}\delta(z-\bar z)\\
    &=-\delta(\bar z-z)\frac{z z_{23}^{2}}{4 \bar z_{23}^{2}  z_{34}^2  z_{21}^{2}}.
\end{align} This factor is exactly the same as the one in \eqref{cpmmp}.
Note that the other parts  can also match with each other once we change the convention. This will be discussed in the graviton scattering.

\paragraph{MHV amplitude for four gravitons.}
 In \cite{Mason:2023mti}, the corresponding Carrollian amplitude is obtained by substituting their eqn.(6.2) into (6.4). Then we could write down the following result in their convention\footnote{We have corrected the eqn. (6.8) in \cite{Mason:2023mti}.}
\bea 
&&\mathcal{C}^{++,--,--,++}(u_1,z_1,\bar{z}_1,\sigma_1;\cdots;u_4,z_4,\bar{z}_4,\sigma_4)\nn\\&=&\kappa_{2,2,-2}^2\frac{\epsilon_1\epsilon_2\epsilon_3\epsilon_4}{(2\pi)^4}\Theta(-Z\Big|\frac{z_{24}}{z_{14}}\Big|^2\epsilon_1\epsilon_4)\Theta(\frac{1-Z}{Z}\Big|\frac{z_{34}}{z_{23}}\Big|^2\epsilon_2\epsilon_4)\Theta(-\frac{1}{1-Z}\Big|\frac{z_{14}}{z_{13}}\Big|^2\epsilon_3\epsilon_4)\frac{1}{\Xi^2}\nn\\ &&\times \frac{1}{Z^2}\delta(\bar Z-Z)\frac{z_{23}^3\bar{z}_{14}^4 z_{34}\bar z_{34}}{\bar z_{23}z_{13}^3\bar z_{13}^3 z_{24}^2 \bar{z}_{24}^2}. \label{cppmm}
\eea
where 
\bea 
\Xi=u_4-u_1Z\Big|\frac{z_{24}}{z_{12}}\Big|^2+u_2\frac{1-Z}{Z}\Big|\frac{z_{34}}{z_{23}}\Big|^2-u_3\frac{1}{1-Z}\Big|\frac{z_{14}}{z_{13}}\Big|^2.
\eea 
For $\epsilon_1=\epsilon_2=-,\epsilon_3=\epsilon_4=+$, the product of the $\Theta$ functions is
\be 
\Theta(Z)\Theta(Z-1)=\Theta(\frac{1}{1-z})\Theta(\frac{z}{1-z})=\Theta(z)\Theta(1-z),
\ee which is the same as our result. Now we compute the factor 
\bea 
&&\frac{1}{Z^2}\delta(\bar Z-Z)\frac{z_{23}^3\bar{z}_{14}^4 z_{34}\bar z_{34}}{\bar z_{23}z_{13}^3\bar z_{13}^3 z_{24}^2 \bar{z}_{24}^2}\nn\\&=&(1-z)^4\delta(\bar z-z)\frac{z_{23}^3\bar{z}_{14}^4 z_{34}\bar z_{34}}{\bar z_{23}z_{13}^3\bar z_{13}^3 z_{24}^2 \bar{z}_{24}^2}\nn\\&=&z^4\delta(\bar z-z)\frac{z_{23}^3\bar z_{12}^2\bar z_{34}^3}{\bar z_{23}^5z_{12}^2 z_{34}z_{13}\bar z_{13}}.
\eea This factor matches with the one in \eqref{fourgraviton}.
\section{Ward identities}\label{Ward}
In this appendix, we will discuss the Ward identities of the Carrollian amplitude in general dimensions. We will first discuss spacetime translation. For scalar field theory, we will use 
\be 
u'=u-c\cdot n\quad \Omega'=\Omega
\ee and expand \eqref{stC} to { the}  first infinitesimal order. Then the Ward identity is 
\bea 
\mathcal{L}^\mu_{\text{st}}[n]\langle \prod_{j=1}^n \Sigma(u_j,\Omega_j,\sigma_j)\rangle=0\label{wdst}
\eea where the differential operator $\mathcal{L}_{\text{st}}[n]$ is 
\be 
\mathcal{L}^\mu_{\text{st}}[n]=\sum_{j=1}^n n_j^\mu \frac{\partial}{\partial u_j}.
\ee  Now we will discuss the Ward identity from Lorentz symmetry for scalar theory. We can expand \eqref{LC} to { the} first infinitesimal order 
\bea 
\mathcal{L}^{\text{scalar}}_{\text{LT}}[Y] \langle \prod_{j=1}^n \Sigma(u_j,\Omega_j,\sigma_j)\rangle=0,
\eea where the differential operator $\mathcal{L}^{\text{scalar}}_{\text{LT}}[Y]$ is 
\bea 
\mathcal{L}^{\text{scalar}}_{\text{LT}}[Y]=\sum_{j=1}^n\left(Y^A(\Omega_j)\frac{\partial}{\partial\theta_j^A}+\frac{1}{2}\nabla\cdot Y(\Omega_j)+\frac{u_j}{d-2}\nabla\cdot Y(\Omega_j)\frac{\partial}{\partial u_j}\right) 
\eea with $Y^A$ the CKVs on the unit sphere $S^{d-2}$.

The discussion can be extended to general spin theory. At first, the Ward identity \eqref{wdst} receives no corrections since the transformation laws of the general spin fields and the scalar field are  the same. The Lorentz transformation law should be modified. We use the vector field theory as an example. We expand \eqref{Card} to { the} first infinitesimal order 
\bea 
\mathcal{L}^{\text{vector}}_{\text{LT}}[Y]\langle \prod_{j=1}^n A_{b_j}(u_j,\Omega_j,\sigma_j)\rangle=0,
\eea where 
\bea 
\mathcal{L}^{\text{vector}}_{\text{LT}}[Y]=\sum_{j=1}^n\left( Y^c(\Omega_j)D_c(\Omega_j)\delta_{a_j}^{b_j}+\frac{1}{2}D_cY^c(\Omega_j)\delta_{a_j}^{b_j} +\frac{1}{2}(D_{a_j}Y^{b_j}-D^{b_j}Y_{a_j})+u_j\frac{D_cY^c(\Omega_j)}{d-2}\frac{\partial}{\partial u_j}\delta_{a_j}^{b_j}\right).\nn\\
\eea 
The Ward identity can be extended to general Carrollian amplitude. They can be unified as 
\bea 
\delta_{\bm\xi}\mathcal{C}_{a_1(s_1)\cdots a_n(s_n)}=0,
\eea where $\bm\xi$ is the vector for the Poincar\'e symmetry at the boundary, and  $\delta_{\bm\xi}$ is the variation of the corresponding field. 
\bibliography{refs}

\end{document}